\documentclass[useAMS,usenatbib]{mn2e}
\usepackage{graphicx}
\usepackage{times}
\usepackage{natbib}
\usepackage{setspace}
\usepackage{amssymb}
\usepackage{amsmath}
\newif\ifAMStwofonts
\AMStwofontstrue

\newcommand{\be}{\begin{equation}}
\newcommand{\ee}{\end{equation}}
\newcommand{\ba}{\begin{eqnarray}}
\newcommand{\ea}{\end{eqnarray}}
\newcommand{\brr}{\begin{array}}
\newcommand{\err}{\end{array}}
\newcommand{\bc}{\begin{center}}
\newcommand{\ec}{\end{center}}

\newcommand{\msun}{M$_\odot$}

\newcommand{\mincir}{\raise
  -2.truept\hbox{\rlap{\hbox{$\sim$}}\raise5.truept \hbox{$<$}\ }}
\newcommand{\magcir}{\raise
  -2.truept\hbox{\rlap{\hbox{$\sim$}}\raise5.truept \hbox{$>$}\ }}
\newcommand{\siml}{\raise
  -2.truept\hbox{\rlap{\hbox{$\sim$}}\raise5.truept \hbox{$<$}\ }}
\newcommand{\simg}{\raise
  -2.truept\hbox{\rlap{\hbox{$\sim$}}\raise5.truept \hbox{$>$}\ }}

  \graphicspath{{figures/}}

\title[Properties of barred spiral disks]
{Properties of barred spiral disks in hydrodynamical cosmological simulations}

\author[D. Goz et al.] {
David Goz$^{1}$, Pierluigi Monaco$^{1,2}$, Giuseppe Murante$^{2}$, Anna Curir$^{3}$\\
$^1$ Dipartimento di Fisica - Sezione di Astronomia, Universit\`a di Trieste, via Tiepolo 11, I- 34131 Trieste -- Italy (goz@oats.inaf.it)\\ 
$^2$ INAF, Osservatorio Astronomico di Trieste, Via Tiepolo 11, I-34131 Trieste -- Italy (monaco, murante@oats.inaf.it)\\
$^3$ INAF, Osservatorio Astronomico di Torino, Strada Osservatorio 20, I-10025 Pino Torinese -- Italy (curir@oato.inaf.it)
}

\begin{document}

\maketitle

\label{firstpage}

\begin{abstract}

We present a quantification of the properties of bars in two
N-body+SPH cosmological simulations of spiral galaxies, named GA and
AqC.  The initial conditions were obtained using the zoom-in
technique and represent two dark matter (DM) halos of
$2-3\times10^{12}\ {\rm M}_\odot$, available at two different
resolutions.  The resulting galaxies are presented in the companion
paper of Murante et al. (2014).  We find that the GA galaxy has a bar
of length $8.8$ kpc, present at the two resolution levels even though
with a slightly different strength.
Classical bar signatures (e.g. pattern of streaming motions, 
high $m=2$ Fourier mode with roughly constant phase)
are consistently found at both resolutions.  Though a close encounter
with a merging satellite at $z\sim0.6$ (mass ratio $1:50$) causes a
strong, transient spiral pattern and some heating of the disk, we find
that bar instability is due to  secular process, caused by a low
Toomre parameter $Q\lesssim1$ due to accumulation of mass in the disk.  The
AqC galaxy has a slightly different history: it suffers a similar tidal
disturbance due to a merging satellite at $z\sim0.5$ but with a mass
ratio of $1:32$, that triggers a bar in the high-resolution
simulation, while at low resolution the merging is found to take place at a later time, 
so that both secular evolution and merging are plausible
triggers for bar instability. 

\end{abstract}

\begin{keywords}
galaxies: formation - galaxies: structure - galaxies: kinematics and dynamics - methods:numerical
\end{keywords}

\section{Introduction}
\label{section:intro}

The formation and evolution of structures within the Lambda Cold Dark
Matter ($\Lambda$CDM) cosmological model is a very active and quickly
evolving field.  In this cosmological framework, galaxies form through
cooling and condensation of baryons within dark matter halos
\citep[e.g.,][]{White_1978,Fall_1980}.  The initial conditions are
provided by cosmology, but the level of complexity of the problem is
so high that following the evolution of galaxies is
a great challenge.  It is then convenient to address the problem
using N-body hydrodynamical simulations.  State-of-the-art
hydrodynamical codes for the formation of galaxies include a treatment
of the processes of radiative cooling, star formation, energy
feedback from massive or dying stars, their chemical enrichment and,
in some cases, accretion onto black holes and feedback from the
resulting active galactic nuclei.  Many of these processes take
place on very small scales, compared with those that can be
resolved by the simulation, so it is necessary to include them through
suitable sub-resolution models.  Thanks to recent progress,
simulations are now able to produce galaxies with realistic
morphologies, sizes and gas fractions.
In particular, despite the relatively unsuccessful simulations shown in the Aquila
comparison project \citep{Scannapieco12}, the
challenge of producing a disk galaxy in a Milky Way-sized halo with
quiet merging history has been successfully carried out by several
groups \cite[e.g.][]{Governato_2007,Guedes_2011,Marinacci_2013,Stinson2013,Aumer13}.

One important observational aspect of disk galaxies is the presence of
a bar: about 60 per cent of nearby disk galaxies are barred when
observed in the near-infrared, while this figure lowers when galaxies
are imaged in the optical \citep{Eskridge_2000,Barazza_2008}.  Bars are
believed to play a key role in the secular evolution of galaxy disks,
particularly in the redistribution of angular momentum of baryonic and
dark matter components \citep{Debattista_1996,Debattista_2000}.  The non-circular
motions of bars cause the migration of gas within the corotation
radius towards the galaxy center, where it can give rise to a
starburst or be accreted on a nuclear black hole.  Also, the formation
of a bar is believed to contribute to the formation of a disky or
boxy/peanut bulge, often called pseudobulge
\citep{Kormendy_1982,Kormendy_2004,Athanassoula_2005,Debattista_2006}.

The emergence of bars in simulated galaxy disks has been addressed in
many papers, starting from the pioneering work of \cite{Ostriker73}
where stability of a disk-shaped rotating N-body system was obtained
only in the presence of an extended spherical halo.  The origin of
bars was ascribed to secular instabilities of massive disks
\citep[e.g.][]{Efstathiou_1982} or to tidal interactions and merging
with galaxy satellites, that excite spiral structures or proper bars
\citep[e.g.][]{Noguchi96,Dubinski08}.  \cite{Sellwood98} noticed that disk
stability is influenced by the presence of a soft or hard center
(namely a gently or steeply rising inner rotation curve), even when
dark matter gives a negligible contribution to the inner part of the
rotation curve.

More recent works addressed the effect of halo triaxiality \citep{Curir_1999} and
concentration \citep{Athanassoula_2002} on the growth of a bar in a
disk hosted by an isolated halo with a \cite{Navarro96} profile.  In
these works the initial conditions represent an equilibrium
configuration of a disk embedded in a dark matter halo; this 
setting is suitable
to study bar formation in the absence of further external
perturbations.  Halo triaxiality was reported by \cite{Curir_1999} to
be a trigger of bar formation, while \cite{Athanassoula_2002} found
that the bar strength correlates with halo concentration.

The study of bar formation in cosmological halos has been faced in
two ways.  \cite{Curir_2006} built zoomed Initial Conditions
(hereafter ICs) of DM halos in cosmological volumes, let them evolve
with an N-body code and placed model disks inside the halos that were
present at some specified redshift.  That paper considered purely
stellar disks, while stellar$+$gas disks were presented in
\cite{Curir_2007} and the effect of star formation was considered in
\cite{Curir_2008}.  In these works it was shown that halo triaxiality,
at the level commonly found in simulated DM halos, 
triggers the formation of bars.  The presence of gas leads to the
destruction of the bar after a few dynamical times if the disk gas
fraction is higher than $\sim20$ per cent, but the switching on of star
formation inhibits this destruction.
Different conclusions were reached by \cite{Berentzen06}, who noticed that the
formation of a bar was weaker in halos with higher triaxiality.
\cite{Athanassoula2013} showed the complex influence of halo
triaxiality on the bar strength: at earlier times it triggers an
instability while, at later times when secular evolution takes place, 
it has a stabilizing effect.

The second way of studying bar formation consists in addressing the
emergence of bars in fully cosmological simulations of the formation
of spiral galaxies
\citep{Scannapieco_2012,Kraljic_2012,Okamoto_2013,Guedes_2013,Okamoto_2014}.  The
obvious advantage of this approach, of a much more realistic
representation of gravitational forces within a non idealized DM halo,
is balanced by the difficulty in obtaining a realistic disk galaxy in
this context.  In particular, extended disks with flat rotation curves
are obtained only when efficient feedback from star formation is
present, but too strong feedback can lead to the destruction of the
disk \citep{Scannapieco12}.  This is where the sub-resolution modeling
of stellar feedback becomes crucial.

\cite{Scannapieco_2012}, using the version of GADGET \citep{GADGET2}
described in \cite{Scannapieco06}, found long and strong bars in two
simulated galaxies; in particular the strengths, lengths and
projected density profiles of the bars were found within the range of
values given in the observations of \cite{Gadotti_2011}.
\cite{Kraljic_2012}, using the Eulerian code RAMSES
\citep{Teyssier02}, studied the evolution of galactic bars in a sample
of 33 zoomed-in cosmological halos.  They found that after $z \approx
1$ almost $80\%$ of spirals galaxies host bars and they suggested
that the epoch of bar formation starts from the late ``secular'' phase
and contributes to the growth of pseudobulges, even if the bulge mass
budget remains dominated by the contribution of mergers.  Furthermore,
most of the bars formed at $z \lesssim 1$ persists up to $z = 0$,
while early bars at $z > 1$ often disappear and reform several times.
\cite{Okamoto_2013} analyzed two simulations of disk galaxies with
disky pseudobulges (Sersic index $<2$), one of which presents a strong
bar and a ``boxy bulge''.  They concluded that, at variance with the
standard picture, the main channel of pseudobulge formation is
high-redshift starbursts and not the secular evolution of the disk,
although this contributes to it in a non-negligible way.
\cite{Guedes_2013} used their Eris simulation to investigate the
interplay between a stellar bar and the formation and evolution of a
pseudobulge.  They found, again at variance with the standard picture,
that the bulk of mass in their pseudobulge forms early ($z \sim 4$),
fast ($\sim2$ Gyr) and in situ, starting from a bar instability
triggered by tidal interactions with a passing satellite.  The bar is
destroyed ($z \sim 3$) by several minor mergers, then reforms later
($z \sim 2$), again triggered by tidal interactions, but the
redistribution of angular momentum within stars and gas, driven to the
center by the bar itself at $z \sim 1$, leads to the gradual
breakup of the bar structure. Very recently \cite{Okamoto_2014}
studied the evolution of two bars formed in cosmological Milky Way-sized halos.
As commonly found in idealized simulations, the rotation speed of
the stronger bar was found to decrease with time by transferring its
angular momentum to the dark matter halo, while other behaviours,
such as oscillations of pattern speed, were more peculiar to the
cosmological case.  The weaker bar was found to slow down, 
while its amplitude was staying constant.  
The authors pointed out that the main difference between idealized and
cosmological simulations is the inclusion of energy and mass released
from stellar populations, which leads to a different central density
structure.

In \cite{Murante14} (hereafter paper I) we present simulations of
Milky Way-sized DM halos performed with the Tree-PM$+$SPH code GADGET3
\citep{GADGET2}, where star formation and stellar feedback are treated
with the sub-resolution model, named MUlti-Phase Particle Integrator
(hereafter MUPPI), presented in \cite{Murante10}.  These simulations
follow the chemical enrichment of the Inter-Stellar Medium (ISM) and
its metal-dependent radiative cooling using the model of
\cite{Tornatore07} and \cite{Wiersma09}.  The two sets of initial
conditions for the simulations, the GA set by \cite{Stoehr02} and the
AqC set of \cite{Scannapieco06}, are available at different refinement
levels and they were used to study the stability of results with
resolution.  The resulting galaxies are shown in paper I to resemble
observed spirals in many regards; in particular, the resulting disks
are extended, with small bulges ($B/T\sim0.2$) and flat rotation
curves.  As shown in that paper, at $z=0$ both spiral galaxies hosted
by the two halos show an extended bar and this feature is clearly
visible at two resolutions both in GA and AqC sets.

In this paper we quantify the properties and kinematics of the bars of
the GA and AqC galaxies of paper I.
We find that, in the GA case and during the development of the 
instability ($z\le0.3$), bar properties and time evolution
are very similar at the two resolutions, while in the AqC case some 
differences are noticed that can be ascribed to the different orbits
and timing of a minor merger at the two resolutions.  

This is, to
our knowledge, the first time that,
in a cosmological simulation, a bar instability is
found to develop in such a similar way at different resolutions.
This prompts
us to consider these bars as due to physical and not to numerical
processes and to study the trigger of this instability,
investigating the role of secular evolution and minor mergers.

The paper is
organized as follows.  Section~\ref{section:simulations} gives a brief
account of the simulations used in this paper.
Section~\ref{section:GA} presents a complete quantification of stellar
orbits and bar strength and length in the GA galaxies, addressing the
question of the physical cause of the bar and the time at which it
appears.  Section~\ref{section:AqC} shows results for the AqC
galaxies, highlighting the differences with the GA simulations and
investigating the physical origin of such differences. Finally,
Section~\ref{section:conclusions} gives a summary and conclusions.

\begin{table*}
\begin{center}
\begin{tabular}{lllllllll}
\hline
Simulation & $\epsilon_{\rm Pl}$ (kpc/h) & M$_{gas}$ (M$_{\odot}$/h) & M$_{star}$ (M$_{\odot}$/h) & N$_{star}$ & $r_{200}/10$ (kpc) & M$_{bulge}$ (M$_{\odot}$) & M$_{disk}$ (M$_{\odot}$) & $B/T$ \\
\hline
GA2        & 0.325                     & $3.0 \cdot 10^{5}$     & $7.5 \cdot 10^{4}$     & 1154243   & $29.98$         & $1.4 \cdot 10^{10}$      & $8.8 \cdot 10^{10}$     & 0.20   \\
GA1        & 0.65                      & $2.8 \cdot 10^{6}$     & $7.0 \cdot 10^{5}$     & 146196    & $30.37$         & $2.0 \cdot 10^{10}$      & $1.1 \cdot 10^{11}$     & 0.22   \\
AqC5       & 0.325                     & $3.0 \cdot 10^{5}$     & $7.5 \cdot 10^{4}$     & 803889    & $23.82$         & $1.1 \cdot 10^{10}$      & $5.7 \cdot 10^{10}$     & 0.23   \\
AqC6       & 0.65                      & $2.4 \cdot 10^{6}$     & $6.0 \cdot 10^{5}$     & 111989    & $24.15$         & $1.4 \cdot 10^{10}$      & $6.3 \cdot 10^{10}$     & 0.24   \\
\hline
\end{tabular}
\end{center}
\caption{Basic characteristics of the different simulations. Column 1:
  simulation name; column 2: Plummer-equivalent softening length for
  gravitational forces, fixed in physical coordinates below $z=6$ and
  in comoving coordinates at higher redshift; column 3: mass of the
  gas particles; column 4: mass of the star particles; column
    5: number of star particles within $r_{200}/10$; column 6: virial
  radius $r_{200}/10$; column 7: stellar bulge mass at $z=0$; column
  8: stellar disk mass at $z=0$; column 9: bulge-over-total stellar
  mass ratio at $z=0$; }
\label{table:ICs}
\end{table*}

\begin{figure*}
\centering{
\includegraphics[angle=90,width=0.49\linewidth]{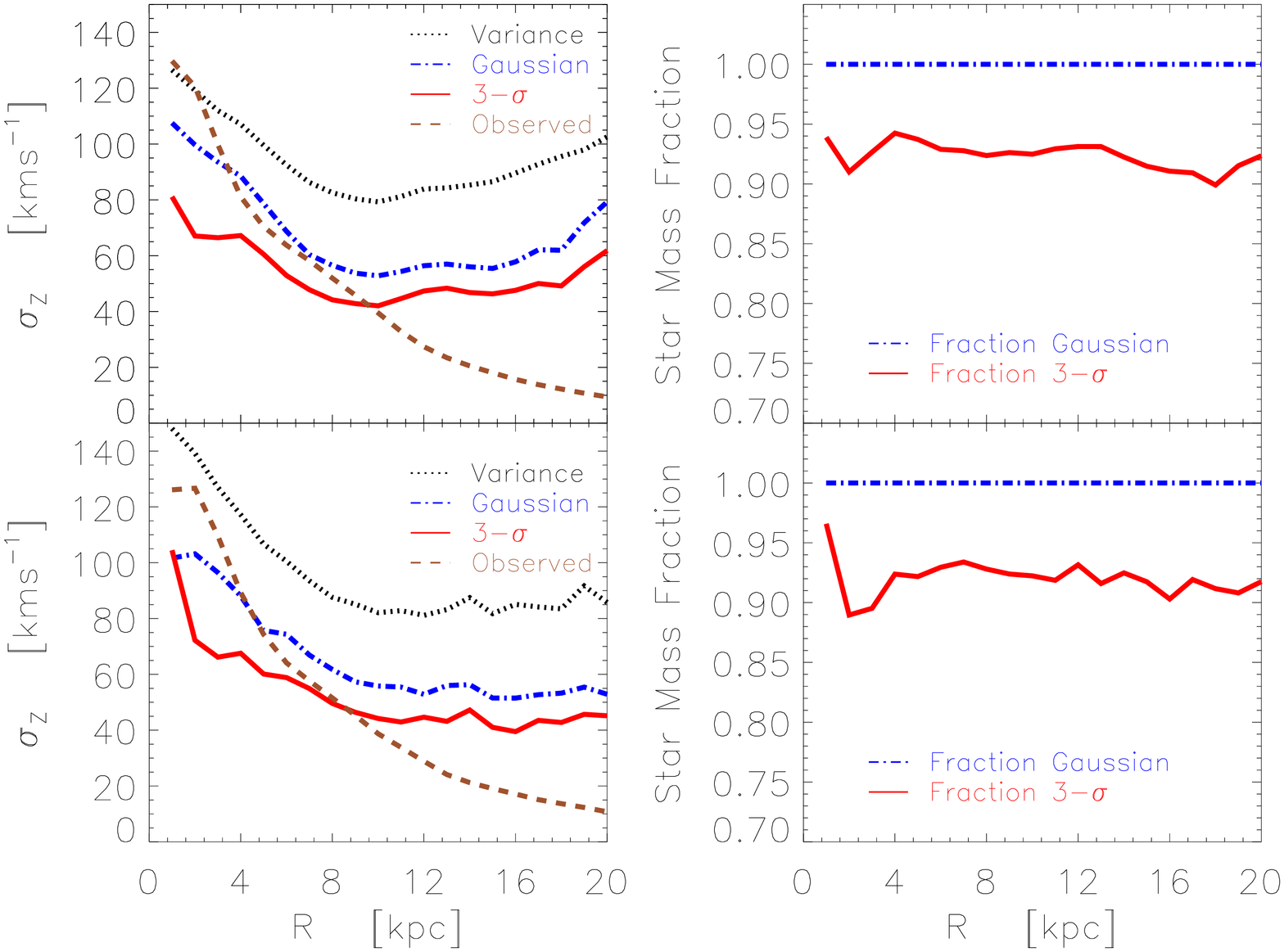}
\includegraphics[angle=90,width=0.49\linewidth]{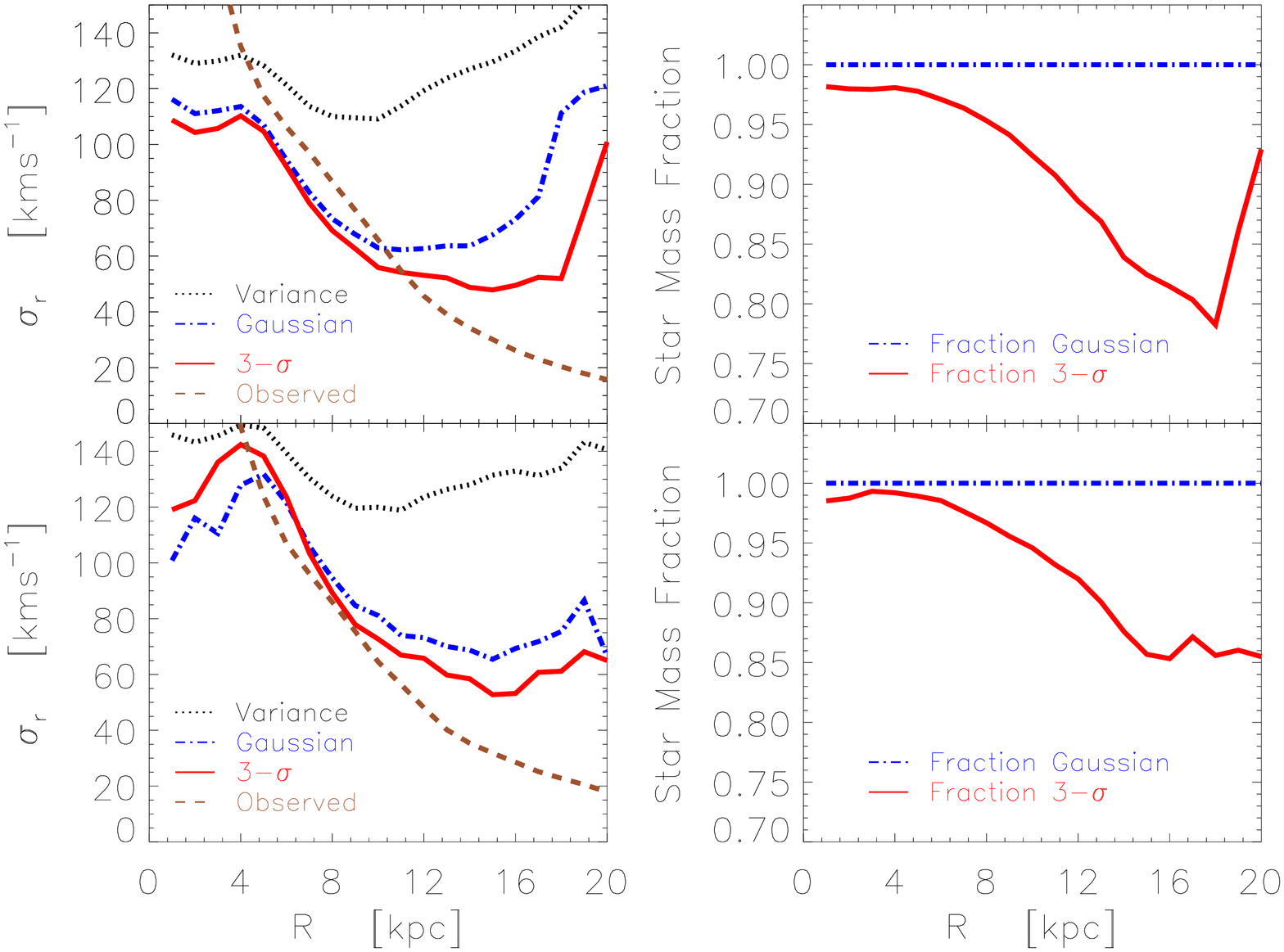}
}
\caption{Vertical velocity dispersion $\sigma_{Z}$ and
  $\sigma_r$, and fractions of mass used to compute the relative
  dispersion; the variance and Gaussian fit methods use all
    particles, so for these the latter quantity is unity, while the $3-\sigma$
    rejection method excludes particles with large velocities and in
    this case the latter quantity gives the fraction of mass that has
    not been excluded.  
  The upper figures give results for GA2, the lower figures for
  GA1.  The left pair of panels gives results for
  $\sigma_Z$, the right pair for $\sigma_r$.  Dotted black, 
  dot-dashed blue, 
  continuous red 
  and dashed brown 
  lines give respectively
  the r.m.s, Gaussian, $3-\sigma$ rejection and observation-oriented
  estimates of velocity dispersion.  }
\label{fig:dispersion}
\end{figure*}

\section{Simulations}
\label{section:simulations}

Simulations were performed using the GADGET3 code \citep{GADGET2},
where gravity is solved with a Tree code, aided by a Particle Mesh
scheme at large scales, while hydrodynamics is integrated with an SPH
solver that uses an explicitly entropy-conserving formulation with
force symmetrization \citep{GADGET2}.  Star formation and feedback are
performed with MUPPI \citep{Murante10}, an effective model that
attempts to take into account the structure of the ISM at unresolved
scales by assuming that each gas particle, at sufficiently high
density, is made up of a cold and a hot phase in thermal pressure
equilibrium, plus a virtual stellar component.  Energy from feedback
is distributed to neighbouring particles both in the form of thermal
and kinetic energy, as described in paper I and in \cite{Murante10}.
Chemical evolution and tracking of 11 elements is performed with the
code of \cite{Tornatore07}, while metal-dependent cooling is performed
following \cite{Wiersma09}.  A full description of the code is given
in the companion paper of paper I, we refer to this paper for all
details.

The two sets of initial conditions that we used are resimulations of
DM halos of mass $\sim3\times10^{12}$ {\msun} (GA) and
$\sim2\times10^{12}$ {\msun} (AqC), with quiet merger history since
$z\sim2$ so as to avoid the risk of late-time major mergers that can
severely damage or destroy the disks.  In particular, the AqC is part
of the Aquarius series \citep{Springel2008} of eight halos, used by
\cite{Scannapieco_2009} to study the formation of disk galaxies;
it was chosen because it was giving the lowest bulge-over-total ($B/T$) ratio 
in that paper. Its level 6 and 5 resolutions were used
in the Aquila comparison project \citep{Scannapieco12}, in which we
participated with an early version of our code with primordial cooling
and purely thermal feedback.  As a matter of fact, both galaxies
happen to suffer a minor merger of mass ratio of order $1:50$ (GA) and
$1:30$ (AqC) at $z\sim0.5$.  For both sets of ICs we use two
resolution levels with initial gas particle masses of
$\sim2\times10^6$ {\msun} (GA1 and AqC6) and $\sim3\times10^5$ {\msun}
(GA2 and AqC5).  Plummer-equivalent softening at $z<6$ is set, for the
two resolutions, to 0.65 and 0.325 kpc/h in physical coordinates, at
higher redshift we keep it fixed in comoving coordinates.

Table\ref{table:ICs} reports the main properties of the four sets of
ICs, together with the main properties of their central galaxies at
$z=0$.
Simulations were post-processed with a standard Friends-of-Friends
(FoF) algorithm to select the main halo of the high-resolution region
and with the substructure-finding code SubFind \citep{Springel01}.  We
assumed that the particles that constitute the galaxy are stars and
cold ($T<10^5$ K) or multi-phase gas particles, hereafter called 
``galaxy particles''. The galaxy was first identified as the
object laying within $1/10$ of the virial radius $r_{200}$, centered
on the center of mass of the FoF halo, then the galaxy position 
was refined by computing the center of mass of galaxy particles
lying within 8 kpc (a distance at which it is unlikely to find
satellite galaxies) and by iterating the computation with the new
center until convergence within 1 pc was reached.  We checked that
this position is very similar to the center of the main substructure
of the FoF halo, computed by SubFind using the position of the most bound particle.
The reference frame was then aligned with the inertia tensor of galaxy
particles, with the $Z$-axis\footnote{
In this paper we use the small letter $z$ to denote redshift and the capital letter $Z$ to denote the spatial vertical axis.}
along the eigenvector corresponding to the
largest eigenvalue and in the direction so as to have a positive
scalar product with the angular momentum; the other two axes were
aligned with the other eigenvectors so as to preserve the property
$\hat{X} \times \hat{Y} = \hat{Z}$.  We checked that, whenever a
significant disk is present, the angular momentum of the
galaxy particles within $r_{200}/10$ is always very well aligned,
within less than one degree, with the $Z$-axis.

\section{The GA1 and GA2 galaxies}
\label{section:GA}

In this Section, we study the GA galaxies obtained
at the two resolution levels, GA1 (lower resolution) and GA2 (higher
resolution). We show both simulations to address the stability of our
result with respect to resolution.

\begin{figure}
\centering{
  \includegraphics[width=0.8\linewidth,angle=90]{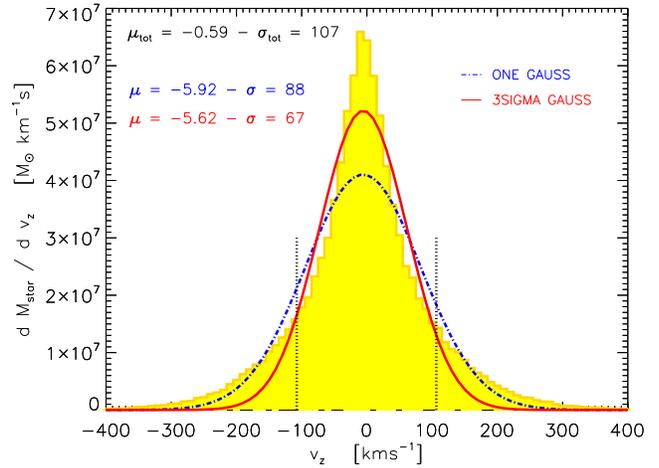}
  }
\caption{Mass-weighted histogram of stellar velocities for GA2 in
  the radial bin from 3 to 4 kpc for the GA2 galaxy at $z=0$.  Measured
  average and r.m.s are $\mu_{tot} = -0.59$ km/s and $\sigma_{tot}
  = 107$ km/s, the r.m.s. is denoted by two dotted black vertical
  lines.  
  The resulting fits (top left of the panel) for
  averages and dispersion with the Gaussian fit (dot-dashed blue 
  line) 
  and 3-$\sigma$ rejection (continuous red 
  line) methods are reported in the figure with the
  same color as the corresponding line, in km/s.   }
\label{fig:velhist}
\end{figure}

\subsection{The vertical structure of the disk}
\label{section:vertical}

The relatively high values of the force softening used in these
simulations do not allow us to resolve the vertical structure of the
stellar thin disk. Nevertheless the kinematic state of the disk is
relevant in the study of disk instabilities, as it is well known that
a hot disk is more stable (see, e.g., the analysis based on the Toomre
parameter presented below). From this point of view, it is a better
choice to address the distribution of stellar velocities in place of
stellar positions. Indeed, the force exhorted by a planar distribution
of mass depends mostly on its mass surface density, a quantity that is
independent of disk scale height, so we expect the velocity dispersion
of stars to be more convergent with resolution than the scale height
itself.  Below, we compute the velocity dispersion both in the
vertical, $Z$-direction and in the radial, $r$-direction.
The latter enters in the Toomre stability criterion.

To compute stellar velocity dispersion, we use all stars within
$r_{200}/10$ from the center, so at each polar radius 
we will have a mix of stars from the disk and from the stellar 
halo\footnote{
We may have some contribution from a thick
disk component, whose study is however hampered by the relatively poor
softening that we use and is anyway beyond the interests of the
present paper.}.
Assuming that the thin disk is the most massive
component and that its vertical or
radial velocities are roughly Gaussian distributed, 
the (mass-weighted) variance of disk stars will be
severely affected by the minor component with a higher velocity
dispersion.  Figure~\ref{fig:dispersion} shows, for both GA2 (upper
panels) and GA1 (lower panels), the mass-weighted root mean square
(hereafter r.m.s.) of vertical (left panels) and radial (right panels) 
velocity dispersion of all stars as black 
lines.
Figure~\ref{fig:velhist} reports, for the radial bin from 3 to 4 kpc,
the mass-weighted histogram of vertical velocities in which black 
vertical lines mark the r.m.s. of the distribution, that is clearly
not representative of the width of the main component.

To improve this measure we tested two options: a Gaussian fit of the
distribution of velocities in each bin of polar radius 
(blue lines in the two Figures) and a Gaussian fit with recursive
rejection of $>3-\sigma$ interlopers, performed until convergence is
reached 
(red lines in the two Figures).  Since the
$3-\sigma$ rejection method implies that some mass is discarded, in
Figure~\ref{fig:dispersion} we report on the right of velocity
profiles the fraction of mass that is used to compute them, as a
function of radial distance. The Gaussian fit method uses all the
mass, so it is reported as a line at unity.  It can be noticed that,
for both velocity components, the Gaussian fit method gives
significantly lower velocity dispersion, while a further suppression
is obtained with the $3-\sigma$ rejection method.  The predicted
distributions are shown in Figure~\ref{fig:velhist}, again as
blue and 
red lines.  From this figure, it is
apparent that the $3-\sigma$ rejection method gives the most faithful
representation of the width of the main mass component, at the modest
cost of excluding less than 10 per cent of mass in the case of
vertical velocity dispersion $\sigma_Z$ and an amount ranging from
1-2 per cent at small radii to $\sim$20 per cent at large distances in
the case of the radial velocity dispersion $\sigma_r$.  We also tried
to fit the velocity distributions with two Gaussians, but the results
were unstable and not satisfactory and for this reason we dropped this method.

In Figure~\ref{fig:dispersion} we also compare these estimates of
$\sigma_Z$ and $\sigma_r$ with observation-oriented estimates.  We
follow \cite{Leroy08} who computed the Toomre $Q$ parameter for the
THINGS sample of local galaxies. Since direct measurements of
stellar velocity dispersion are {\it ``extremely scarce''}, in their
Appendix B they assumed that the stellar scale height $h_\star$ is
constant throughout the disk, that this same quantity is related to
the disk scale radius $r_\star$ through the observed relation $h_\star
= r_\star / 7.3$ and that stars are isothermal in the $Z$-direction.
As a result:

\begin{equation}
\sigma_{Z} = \sqrt{\Sigma_\star \frac{ 2\pi G r_\star }{ 7.3 }  }\, ,
\end{equation}

\noindent
where $\Sigma_\star$ is the disk stellar surface density.
For the radial velocity dispersion, the same authors assumed that $0.6
\sigma_r=\sigma_Z$.  For the GA1 and GA2 disks we use a disk scale
length $r_{*} = 3.93$ and $4.45$ kpc, estimated in paper I by fitting
the radial profile of stellar mass surface density.  These estimates
are shown in the Figure~\ref{fig:dispersion} as 
brown lines.  As a result, at $r<10$
kpc both the Gaussian and the $3-\sigma$ rejection methods yield
velocity dispersions that are broadly compatible with those that would
be expected in an observed galaxy.  In more detail, GA1 shows a
marginally hotter disk, especially as far as radial motions are
concerned, but these are influenced by the radial streaming motions
due to the bar itself, that will be quantified below.  The outer parts
of the disk are significantly hotter than this estimate, but we know
that, at large distances, disk scale lengths start to increase, so this discrepancy 
could be not very significant.

This analysis demonstrates that, despite the relatively large
softenings used, the kinematic state of our disks is compatible with
observational evidence, at least in the inner regions that are subject
to bar instability.  However, we know that these velocity dispersions
are severely affected by numerics and the marginally hotter GA1 disk
clearly supports this warning.  In paper I, we present results of
simulations of the GA1 obtained with different softenings and we show
that, while lower values of the softening result in higher stellar
velocity dispersion (simply measured there as r.m.s.), convergence is
not yet fully achieved for the value of softening used here. Therefore 
these velocities are very likely influenced by 2-body scattering of star
particles, but this influence does not lead to unrealistic thickening
of the disk.

\begin{figure*}
\centering{
\includegraphics[angle=90,width=\linewidth]{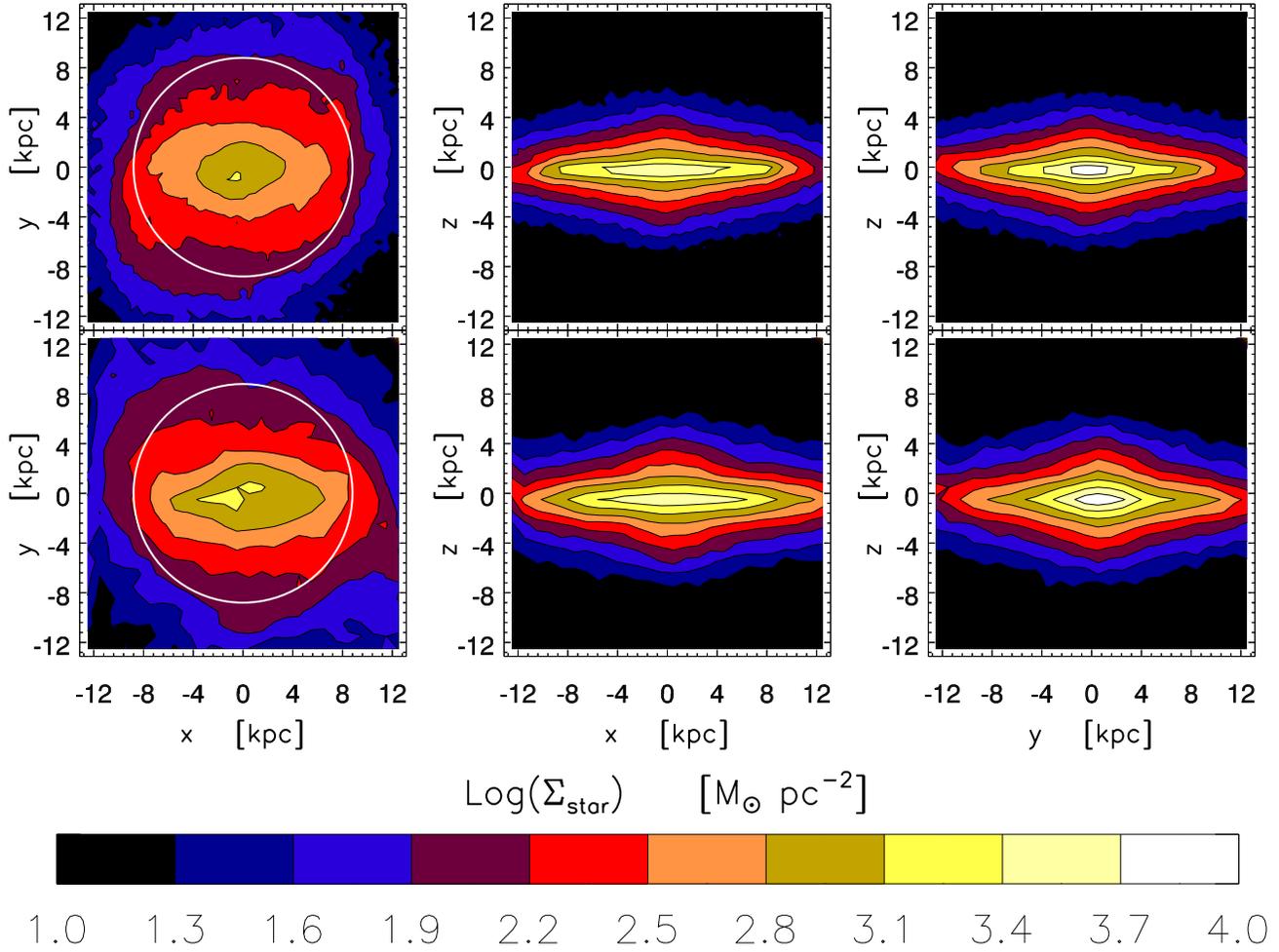}
}
\caption
{Projected stellar maps of the GA2 (upper panels) and GA1 (lower
  panels) galaxies at $z=0$ in boxes centered on the galaxy that
  extend 25 kpc across. Color coding follows the Log of stellar mass
  surface density as indicated by the color bar. The three columns
  show projections in the face-on $XY$-plane (left), edge-on
  $XZ$-plane (middle) and edge-on $YZ$-plane (right).  The white
  circle marks the bar length $L_{\rm bar}$. }
\label{fig:maps}
\end{figure*}

\subsection{Morphology and circularities}
\label{section:morph_GA}

Extended density maps of the two galaxies are given in paper I.
Figure~\ref{fig:maps} shows maps of stellar mass surface density for the
central part of the two simulated galaxies at redshift 0. Each map
spans $\pm12.5$ kpc in each dimension.  The figure shows, on the upper
panels, the maps of the GA2 simulation in the $XY$-, $XZ$- and
$YZ$-plane, while the lower panels show the same maps for the GA1
simulation.  The color coding represents the Log surface density and
values are given in the color bar.  The maps were obtained by
projecting all star particles within $r_{200}/10$ and smoothing the
resulting surface densities on a grid, whose pixel is set equal to the
softening length in both cases.  We find that in both simulations the
isodensity contours in the face-on map are not round but present a
flattened structure; the bar is aligned along the $X$-axis.
The edge-on maps sample the bar along its long 
($XZ$ projection) and short ($YZ$ projection) axes and,
as expected, the isodensity contours are flatter
when the disk particle distribution is seen along the major bar axis.

\begin{figure*}
  \centering{
  \includegraphics[angle=90,width=\linewidth]{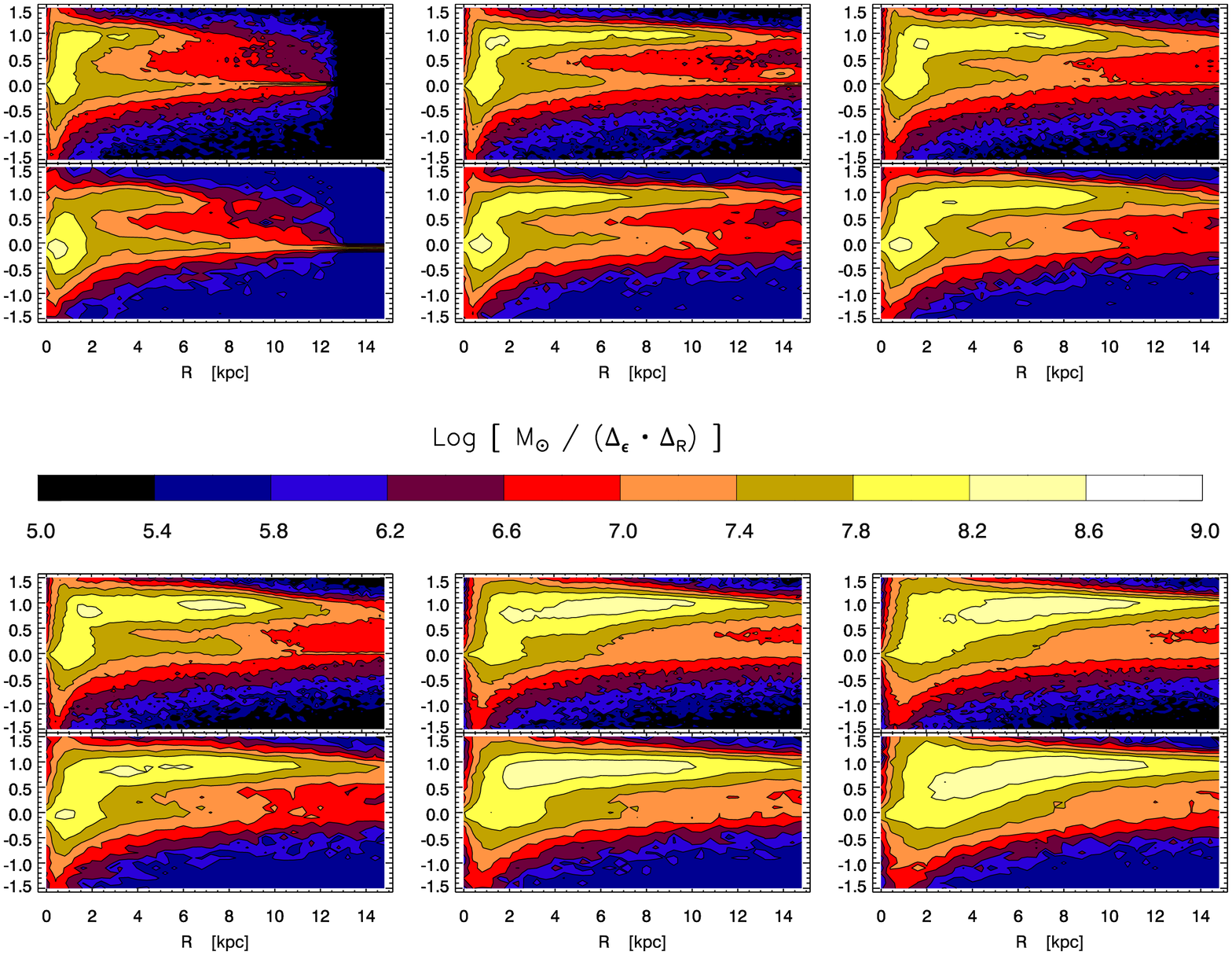}
  }
\caption{2D histograms of circularity (Eq.~\ref{eq:circ}) and polar
  radius, at six different times.  In each panel, color coding refers
  to the Log of stellar mass per unit bin area as indicated by the
  color bar, the black lines are
  the corresponding isodensity contours.  Each pair of panels gives
  results for the GA2 (upper panels) and GA1 (lower panels) galaxies.
  From the upper left panels, redshifts are $z =1.5$, $1$, $0.7$, $0.5$, $0.2$,
  $0$.}
\label{fig:circ}
\end{figure*}

The circularity of a star particle is usually defined as the ratio of
the $z$-component of its specific angular momentum, $j_z=r v_{\rm
  tan}$ (where $v_{\rm tan}$ is the tangential velocity in a
cylindrical coordinate system) and the angular momentum of a
reference circular orbit.  Two methods have been proposed to compute
this reference angular momentum.  \cite{Scannapieco_2009} and other
authors used the angular momentum of a circular orbit at the same
radius, $j_{\rm circ}=r \sqrt{GM(<r)/r}$, while \cite{Abadi_2003}
proposed to use $J_{\rm max}(E)$, the maximum specific angular momentum possible
given the binding energy of the particle $E$. This second
definition constrains circularity to be $<1$.
As in paper I, in this paper we use the first definition, so that:

\be \epsilon = j_Z/j_{\rm circ} \label{eq:circ} \ee 

\noindent
However, we also tested the second definition at $z=0$ and found 
the same qualitative features of the circularity diagrams obtained
with Equation~\ref{eq:circ}.

In Figure~\ref{fig:circ} we show  2D histograms of circularities as a
function of polar radius $r$ along the galaxy disks, for both
galaxies at six different times. We expect a disk to be visible as a
narrow distribution around $\epsilon \sim 1$, while a component scattering
around $\epsilon \sim 0$ will be identified as a bulge or a spherical halo,
depending on $r$.
The 2D histograms are shown as a map of mass per
unit bin size (kpc and circularity in the two dimensions), with
isodensity contours showing the preferential locus of star particles.
For each row, the upper and lower panels give the histogram
for GA2 and GA1 respectively.  Starting from the upper left panels,
the two rows show results for $z=1.5$, $1$, $0.7$, $0.5$, $0.2$, $0$.
These times are chosen to follow the main phases of the formation of
the bar, as it will be explained below. This figure allows us to monitor
the formation of the disk in the two simulations. The first point
that is worth noting is that the circularity histograms are notably
independent of resolution, so we will describe GA1 and GA2 together.
At $z=1.5$ the galaxy is mostly a spheroid, while a disk has started
to form at $r<6 kpc$. A clear and thin disk structure is visible at $z=1$.  The
structure is broader at $z<0.6$, where tidal interactions with a
satellite of mass $M_\star=1.2\times10^9$ {\msun}, 
that culminate with a minor merger at $z=0.35$, are heating the
disk.
Beyond $z=0.2$, the region at intermediate circularities,
$\epsilon\sim 0.5$ and at $1<r<5$ kpc, starts to be populated,
especially for the GA1 galaxy. Because stars in the bar have
  large systematic radial motions, this is a sign of the emergence of a
bar structure.

The fraction of stellar mass as a function of circularity, i.e
  the projection in radius of the circularity histograms of
  Figure~\ref{fig:circ}, is shown in figure 3 of paper I for $z=0$.
  There a simple, kinematically-based decomposition \citep[that
    loosely follows][]{Scannapieco08} is used to compute the stellar
  $B/T$.  We assume that counter-rotating star particles ($\epsilon <
  0$) within $r_{200}/10$ constitute half of the spheroidal component,
  so the bulge mass is computed as twice the mass of these
  counter-rotating stars.  $B/T$ was then computed as:

\be B/T = M_{\rm bulge} / ( M_{\rm bulge} + M_{\rm disk}
) \label{eq:BT}\ee

\noindent
$B/T$ values at $z=0$ are reported in Table~\ref{table:ICs}.

\subsection{Kinematics of the central region}
\label{section:bulge}

Here we quantify the kinematics of the central region of the galaxy,
that hosts both bar and bulge.  It will be demonstrated in
Section~\ref{section:birth} that in GA1 as well as in GA2 a bar appears
only towards the end of the simulation, starting from $z=0.2$, so we
analyze here only the galaxy at $z=0$.

\begin{figure}
\centering{
\includegraphics[angle=90,width=\linewidth]{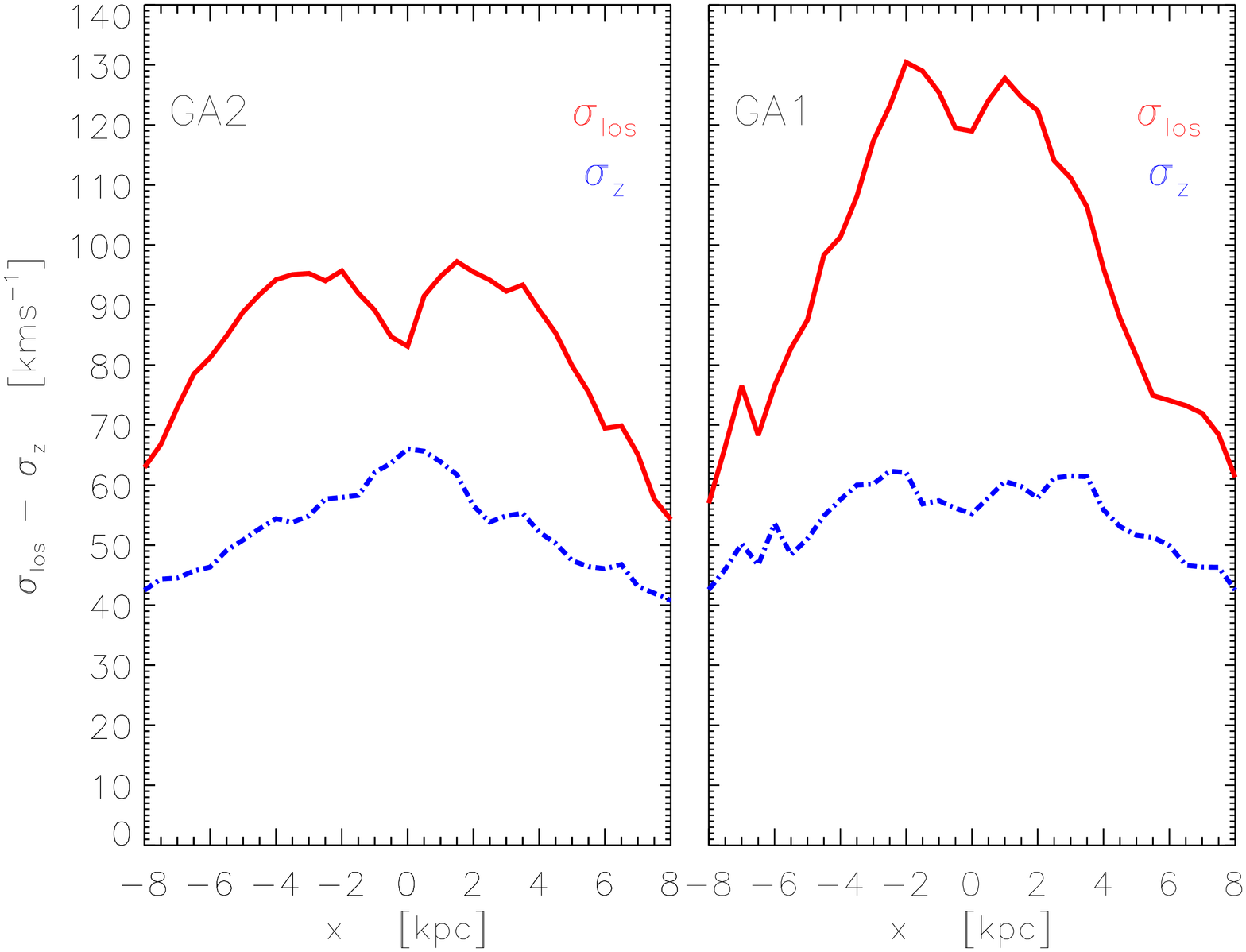}
}
\caption{Profiles of line-of-sight velocity dispersion $\sigma_{los}$ (continuous red
  lines) and vertical velocity dispersion $\sigma_Z$ (dot-dashed blue lines) of
  stars at $z=0$, when the galaxy is observed edge-on in the direction
  perpendicular to the bar main axis. Left and right panels show GA2
  and GA1.}
\label{fig:v_dispersion}
\end{figure}

As suggested by \cite{Okamoto_2013}, the expected kinematic signature of a bar is a
higher value of line-of-sight velocities with respect to vertical
ones. 
We repeat the analysis suggested in that paper in 
Figure~\ref{fig:v_dispersion}, that shows velocity dispersion $\sigma_{los}$ of
stars along the ``line of sight'' (the $Y$ axis in the density maps;
red lines in the figure) and along the vertical direction $\sigma_Z$ ($Z$ axis;
blue lines in the figure).  Velocity dispersions were computed 
with the 3-$\sigma$ rejection method explained in Section \ref{section:vertical}.

We find that, while vertical velocity dispersion gets values of $\sim50-60$ km/s
with a very flat dependence on distance from the center, radial
velocity dispersion takes higher values.  In the GA2 case, radial
velocities peak at $\sim100$ km/s, twice the vertical
ones, while in the GA1 case an even higher peak ($130$ km/s) is
present.  In both cases, a small dip at the galaxy center is present.

\begin{figure}
\centering{
\includegraphics[angle=90,width=\linewidth]{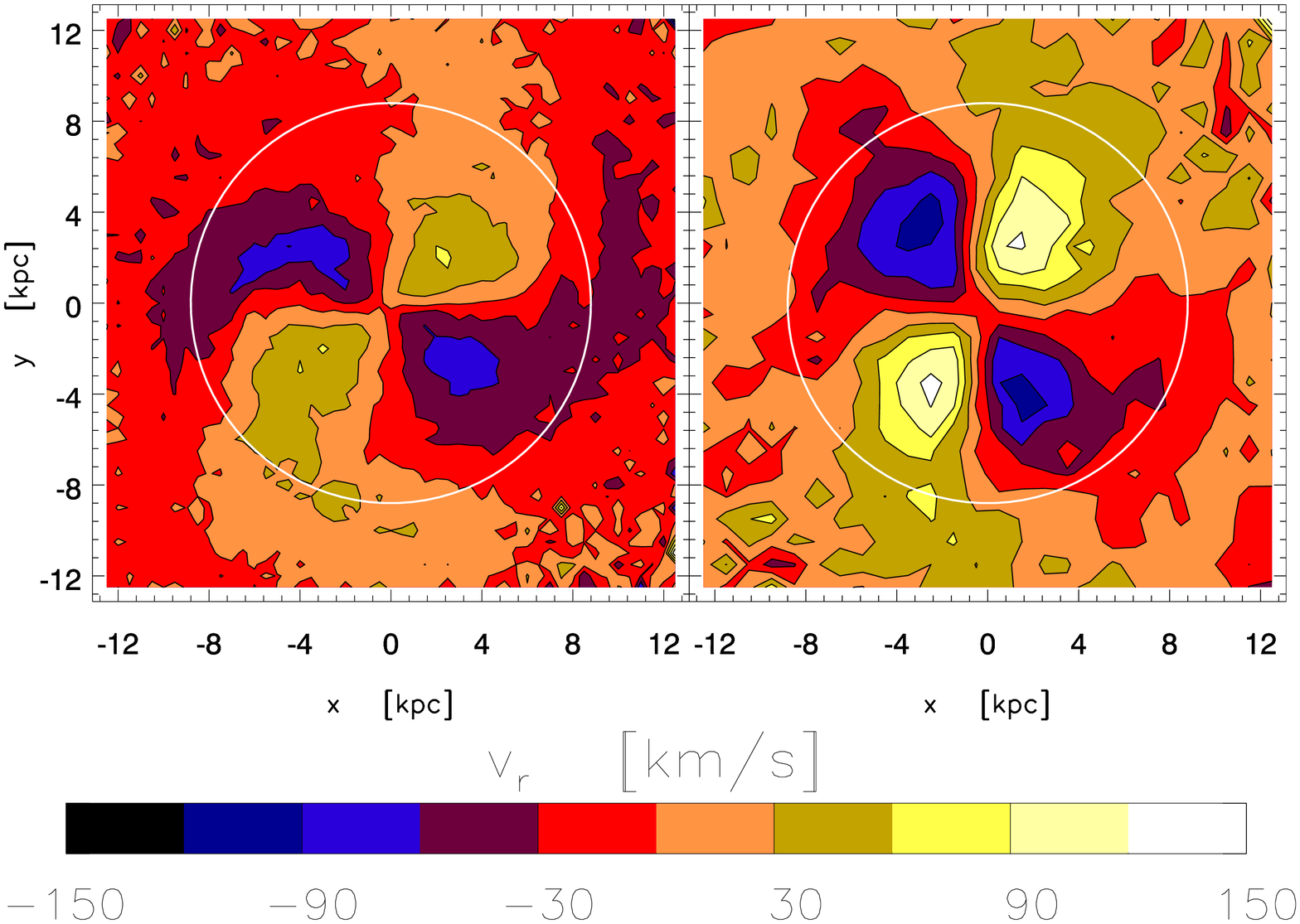}
}

\caption{Radial velocity maps on face-on views ($XY$ projection) of
  the two simulated GA2 (left panel) and GA1 (right panel) galaxies.
  The boxes are 25 kpc across.  Color coding follows the median radial
  velocity of stars within 1 kpc from the midplane, as indicated by
  the color bar.  The white circle marks the bar length $L_{\rm bar}$.}
\label{fig:streaming_motion}
\end{figure}

A pronounced, radial streaming pattern is expected in
non-axisymmetric potentials like that of a stellar bar 
\citep[e.g.][]{Bosma_1978}.
Figure~\ref{fig:streaming_motion} shows 2D maps, in the face-on 
$XY$ plane, of average radial velocities computed on the
same box size and grid as in Figure~\ref{fig:maps}. To minimize
contamination from halo motions, median velocities
are computed only for stars lying within 1 kpc from the midplane.  As
shown in the color bar below, blue/black colors denote
negative velocities, while yellow/while colors denote positive
velocities. A symmetric and squared pattern of streaming motions
is evident in both cases,
with higher velocities for the GA1 galaxy. At larger distances
from the center, the
velocity pattern connected to the spiral arms is very visible for GA2.  We
conclude that the kinematics of stars in the inner regions of these
galaxies is dominated by streaming motions as expected.

\subsection{Quantification of bar strength and length}
\label{section:strenght}

We quantify the strength of the bar following the analysis of
\cite{Scannapieco_2012} and \cite{Kraljic_2012}. This is based on the Fourier transform
of the surface density of the disk:

\begin{equation}
 \Sigma \left(r,\theta \right) = \frac{a_{0}(r)}{2} + \sum_{n=1}^{\infty}\left[ a_{n}(r)\cos(n\theta) + b_{n}(r)\sin(n\theta)\right]
\end{equation}

\noindent
where $r$ is the polar radius and $\theta$ the azimuthal position on the
disk plane. To perform the transform, particles are radially binned
and the following coefficients are computed for each bin:

\begin{eqnarray}
  a_{n}(r) &=& \frac{1}{\pi}\int_{0}^{2\pi} \Sigma(r)\cos(n\theta)d\theta\,,  \left(n\geq 0\right) \label{eq:fourier1} \\
  b_{n}(r) &=& \frac{1}{\pi}\int_{0}^{2\pi} \Sigma(r)\sin(n\theta)d\theta\,,  \left(n > 0\right)   \label{eq:fourier2}
\end{eqnarray}

\noindent
where $\Sigma$ is the surface density of the stellar disk. 
The Fourier amplitude of each mode is defined as:

\begin{equation}
 C_{n}(r) = \sqrt{a_{n}^{2}(r) + b_{n}^{2}(r)}\,,\ \ \ \  C_{0}(r) = a_{0}(r)/2
\end{equation}

\noindent
The presence of a bar is revealed by a significant value of the $C_2$ component,
that is higher than even components of further order.

\begin{figure}
\centering{
  \includegraphics[angle=90,width=\linewidth]{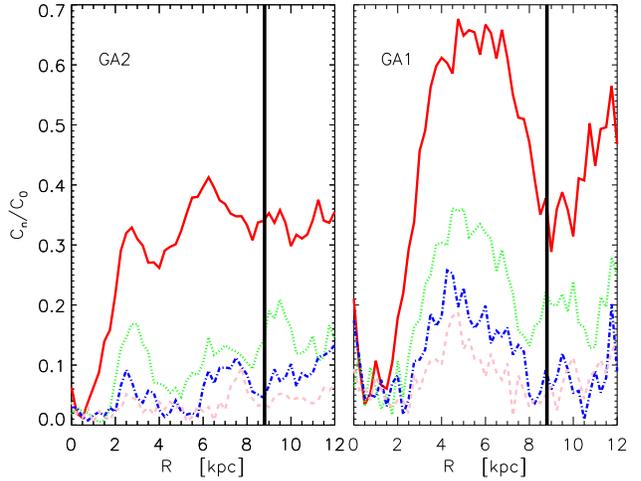}
  }
\caption{Fourier amplitudes $C_n/C_0$ for even components $n=2$ (continuous red
  line), $n=4$ (dotted green line), $n=6$ (dot-dashed blue line), $n=8$ (dashed pink line) for
  stars of GA2 (left panel) and GA1 (right panel). The black lines
  mark the bar length $L_{\rm bar}$ kpc.}
\label{fig:fourier}
\end{figure}

In Figure~\ref{fig:fourier} we show the amplitudes of the first four
even components, normalized to the $n=0$ one, $C_{n}/C_{0}$.  In both
cases the $C_2$ component is significantly higher than the other
components; for the GA2 it peaks at a value of 0.4 at 6 kpc, with a
broad plateau starting from 2 kpc.  GA1 shows a narrower plateau
between 5 and 7 kpc, reaching a higher value of 0.7.  This confirms
that, consistently with the higher radial velocities, the bar in the
GA1 simulation is stronger than in GA2.  Higher order moments show
smaller and smaller values in both cases; they peak at different
scales and this is a sign that the bar is not perfectly symmetric.  
We also checked that odd modes have small values and this is
again consistent with what we expect from a bar.

These results are broadly consistent with observations that show a
variety of radial Fourier profiles of bars, ranging from simple
symmetric profiles, that can be represented by two overlapping Gaussian
components, to more complex curves. Since
$C_{n}/C_{0}$ value spans between 0.4 and 0.8
\citep{Elmegreen_1985,Ohta_1990,Ohta_1996,Aguerri_1998,Aguerri_2003,Buta_2006}, 
both GA galaxies would be classified as barred. 
Moreover, the amplitudes of
GA2 show relatively high values also at large radii, where the
signature of streaming motions (Figure~\ref{fig:streaming_motion}) is already
lost but prominent spiral arms are present. This shows a limit of the
analysis based simply on Fourier amplitudes, where spiral arms give
weak signatures that can be confused with those of bars.

\begin{figure}
\centering{
  \includegraphics[angle=90,width=\linewidth]{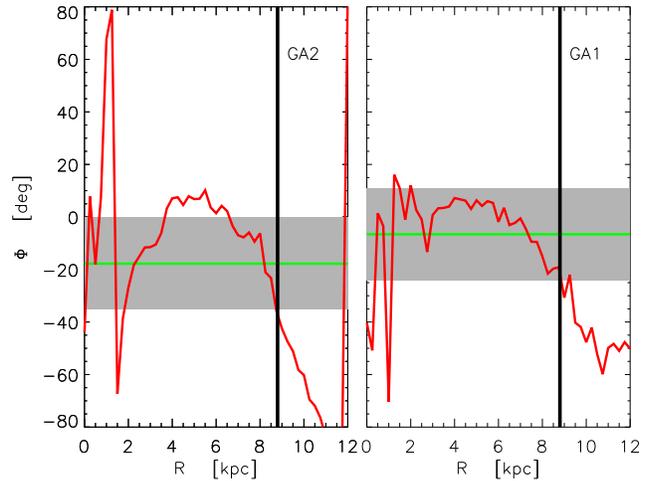}
  }
\caption{Phase of stellar distribution for GA2 (left panel) and GA1
  (right panel). Red lines report the phase $\Phi$ of the $n = 2$ mode
  as a function of radius, the green horizontal line marks the overall
  phase $\Phi_{\rm disk}$ of the disk within 12.5 kpc. The shaded
  area gives the allowed range of $\Phi_{\rm disk} \pm \arcsin (0.3)$.
  The black lines correspond to the bar length $L_{\rm bar}$.}
\label{fig:phase}
\end{figure}

To evaluate the length of the bar, among many published methods
\cite[e.g.][]{Combes_1993,Debattista_2000, Michel-Dansac_2006}, we
use the one proposed by \cite{Athanassoula_2002}.   This is based on
the fact that, for an ideal bar, the phase of the $m=2$ mode should be
constant as long as the probed scale is within the bar, while beyond
it the phase is expected to fluctuate due to spiral arms.  
We implement this method by calculating, for each radial bin, the
phase $\Phi$ of the $m=2$ mode as:

\begin{equation}
\Phi(r) = \arctan \left( b_2(r)/a_2(r) \right) \label{eq:phase}
\end{equation}

\noindent
The average phase is computed applying Equations~\ref{eq:fourier1}
and \ref{eq:fourier2} to all star particles with $r<12.5$ kpc, then
computing $\Phi_{\rm disk}$ as in Equation~\ref{eq:phase}.  The result
is reported in Figure~\ref{fig:phase} as a red line, while the green
one corresponds to the average phase.  The bar length is defined as
the largest radius where these two last quantities differ less than
a certain value. 
As discussed in \cite{Athanassoula_2002}, the choice of the
constant is somehow arbitrary. In that paper the authors suggested 
a range of $\pm \arcsin(0.3)$, while tighter ranges were used by other authors
(e.g. \citet{Kraljic_2012} used $\pm \arcsin(0.1)$).  We will
demonstrate below that the GA2 bar is a caught in an early
development phase and it is still weak, so we adopt the more
permissive criterion of $\pm \arcsin(0.3)$.
For both GA2 (left panel) and GA1 (right 
panel) the phase fluctuates in the inner 1-2 kpc, then is relatively
stable for several kpc and goes out of the shaded region at 8.8 kpc in
both cases.  It must be noticed that the phase of GA2 gets marginally
above the shaded region from 4 to 6 kpc, but given the arbitrariness
of the used value we neglect this minor issue.  We then take $L_{\rm
  bar}=8.8$ kpc as a measure of bar length (see the vertical black
line in the Figure) and notice that it is remarkably independent of
resolution.

Another possible method proposed by \cite{Athanassoula_2002} to
evaluate bar length is based on the scale at which the Fourier
coefficient $C_2/C_0$ goes to zero. Indeed, for an ideal bar on an
axisymmetric disk, we would expect this coefficient to show a plateau
and then drop quickly beyond the bar. In a more realistic context one
should define a threshold with respect to the maximum and define the
bar length as the radius at which the amplitude of the Fourier mode gets
below it.  In Figure~\ref{fig:fourier} the vertical black lines denote
the bar length as estimated by the phase method.  In the GA1 case,
using a drop of the coefficient by a factor of two, it  would give almost 
the same bar length, while, as noticed above, in the GA2 case the spiral
pattern gives a signal comparable to that of the bar.  
$L_{\rm bar}$ is reported as a circle also in Figures~\ref{fig:maps}
and \ref{fig:streaming_motion} and in both cases the estimated bar length
separates the inner region, dominated by flattened isodensity contours
and streaming motions, from the outer region dominated by spiral arms.  We
conclude that the phase method gives a fair estimate of $L_{\rm bar}$.

We can compare our results with $\sim 300$ observed galaxies presented
in \cite{Gadotti_2011}; our $L_{\rm bar}$ value is at the high end of
the distribution given in this observational work, but well compatible
with it.

\subsection{The origin of bar instability}
\label{section:birth}

\begin{figure}
\centering{
\includegraphics[angle=90,width=\linewidth]{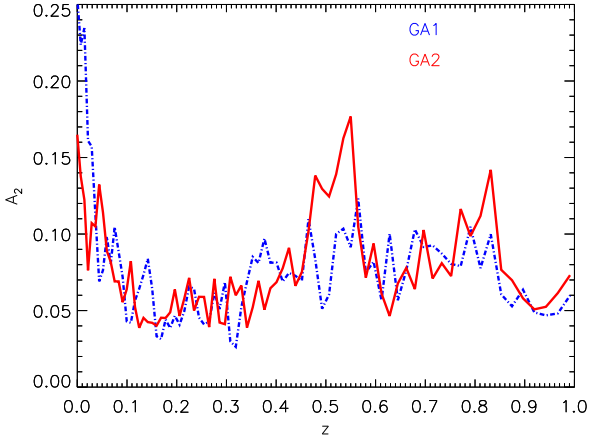}
}
\caption{Relative mode strength $A_{2}$ as a function of redshift for
  the GA1 (dot-dashed blue) and GA2 (continuous red) galaxies, starting from $z=1$.}
\label{fig:amplitude}
\end{figure}

To estimate the time at which the bar is triggered, we quantify how
bar strength grows by computing the so-called relative mode strength
$A_{2}$.  This is the ratio of the integrals, weighted by area, of the
coefficients $C_{2}$ and $C_0$ over the bar length $L_{\rm bar}$,
taking into account its value estimated by the previous method in 
Section \ref{section:strenght}.

\begin{equation}
 A_2 = \frac{\int_{0}^{L_{\rm bar}}C_{2}(R)R dR}{\int_{0}^{L_{\rm bar}}C_{0}(R)R dR}
\end{equation}

\begin{figure*}
\centering{
\includegraphics[angle=90,width=0.49\linewidth]{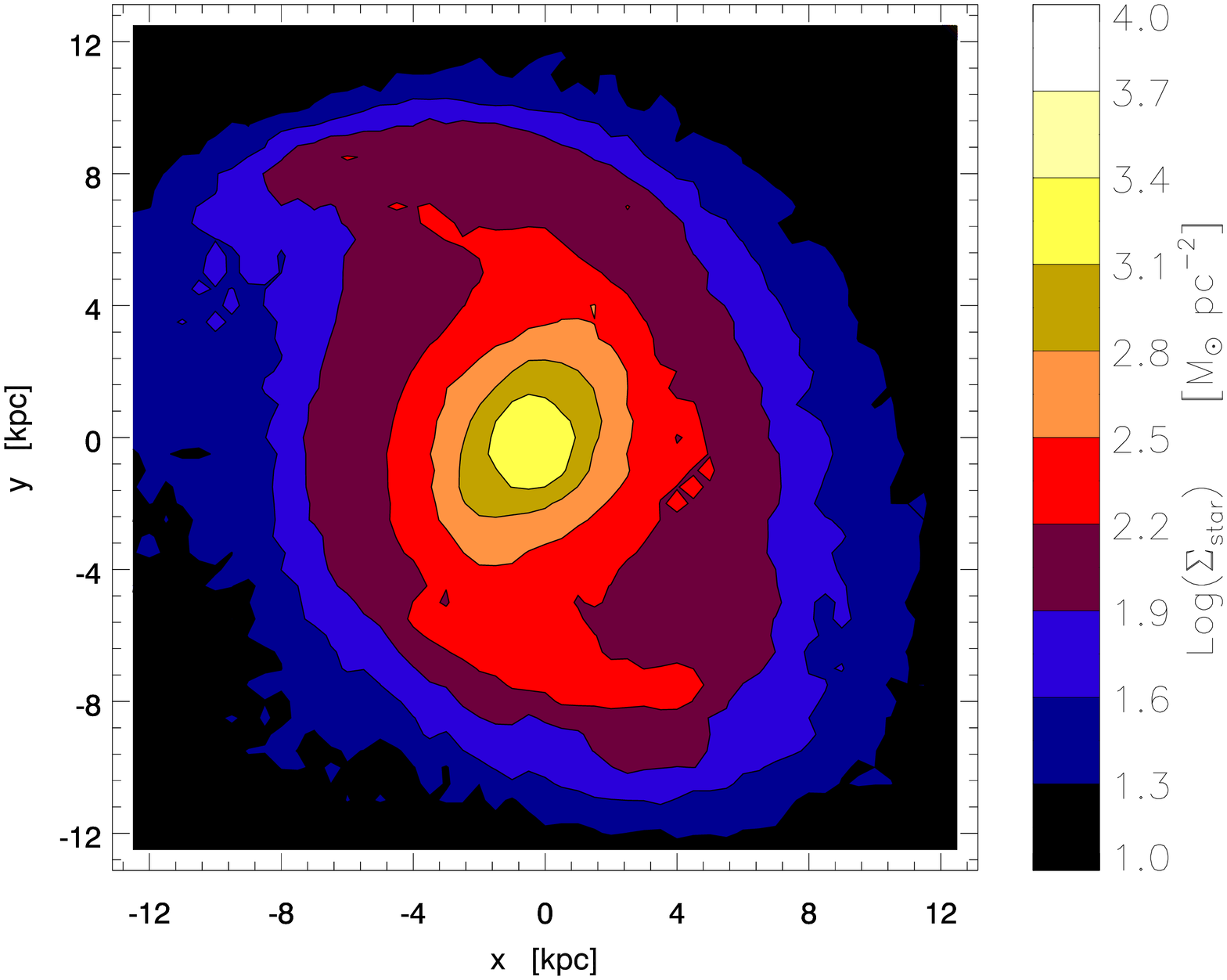}
\includegraphics[angle=90,width=0.49\linewidth]{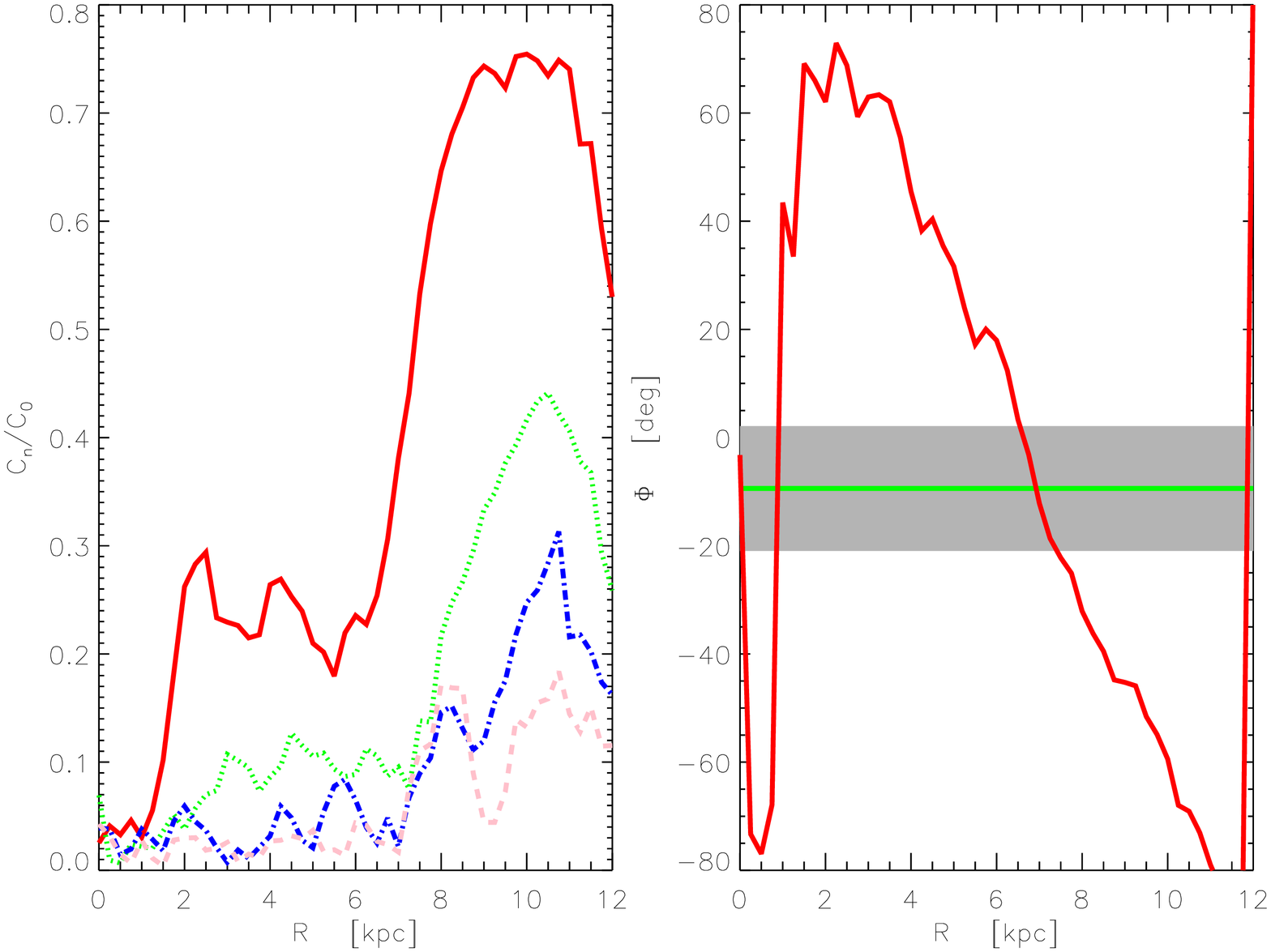}
}
\caption{Stellar mass surface density maps (left panels), Fourier
  amplitudes (middle panel) and phase (right panel) for
  the GA2 galaxy at $z=0.55$.  Symbols and colors are like in
  Figures~\ref{fig:maps}, \ref{fig:fourier} and \ref{fig:phase}.}
\label{fig:minormerger}
\end{figure*}

\noindent
We perform this calculation for all simulation outputs since $z=1$,
when the disk is still young and both GA1 and GA2 show no sign of a
bar.  Figure~\ref{fig:amplitude} shows the evolution of $A_{2}$ with
redshift for GA2 (red line) and GA1 (blue line).  At $z<0.7$, two
different episodes of growth of $A_2$ are visible at $z$ from $0.6$ to
$0.45$ (at least for GA2) and at $z<0.2$.  The episode at $z\sim0.5$
is due to a minor merger already mentioned above.  A
satellite of stellar mass $1.25\times10^9$ {\msun} in GA2 and
$1.35\times10^9$ {\msun} in GA1 performs close orbits around the main
galaxy. For GA2, the closest encounter is found at $z=0.57$, where the
distance of the satellite from the galaxy center is 12 kpc.  The
stellar mass ratio is $1:50$ at the beginning of the interaction,
slowly decreasing because of the continuous growth of the stellar mass of
the central galaxy.  Two more close encounters are found before the
satellite is tidally disrupted into the halo of the main galaxy ($z=0.35$).
These interactions trigger non-axisymmetric perturbations, so that GA2
acquires $A_2$ values equivalent to those at $z=0$.  During this
period the disk is noticeably disturbed and a bar-like morphology is
visible only in one output (the time interval between outputs being
$\sim100$ Myr).  In Figure~\ref{fig:minormerger} we show in the left
panel a face-on density map of GA2 at $z=0.55$ (just after the nearest
encounter), analogous to Figure~\ref{fig:maps}, and in the right panels
the $C_2/C_0$ and phase diagrams, analogous to
Figures~\ref{fig:fourier} and \ref{fig:phase}.  While disk morphology
is clearly disturbed, the $C_2$ coefficient takes on significant
values especially at large radii, while the phase analysis reveals
that the structure does not behave as a bar.  So the high value of
$A_2$ is determined by the outer spiral arms more than by a central
bar.  This transient lasts from the first near passage of the
satellite to its destruction, $\sim700$ Myr in total, corresponding to
$\sim3-4$ revolutions of the disk.  Then the $A_2$ coefficient quickly
returns to $\sim0.1$ and keeps decreasing slowly with time.

The history of GA1 is similar.  In this case the satellite is never
found at distances smaller than 20 kpc so, while the mass ratio is
very similar, the tidal disturbance is smaller and the jump in $A_2$
is much less evident.

The $A_2$ coefficients start to grow for both galaxies after $z=0.2$,
$\sim2$ Gyr after the minor merger has ended (i.e. after $\sim10$ disk
revolutions).  In this period the
instability grows at an accelerating pace.  The growth has a markedly
different time evolution with respect to the instability episode
triggered by the minor merger. This suggests that the bar is due to the
secular evolution of the disk. Moreover, the nice coincidence of
the timing of bar growth at the two resolutions and the similar length
of the resulting bar suggest that this instability is
physical and not purely numerical, while the difference in bar
strength is explained by the quick raise of the bar instability, so
the GA1 at $z=0$ happens to be caught at a higher $A_2$ value.

\begin{figure}
\centering{
  \includegraphics[angle=90,width=\linewidth]{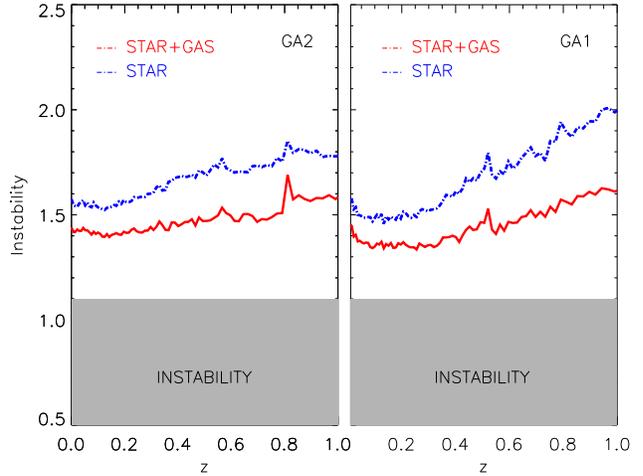}
  }
\caption{$\Upsilon_{disk}$ as a function of redshift for star
  particles (dot-dashed blue line) and for all galaxy particles (continuous red line); the
  shaded area marks the instability region corresponding to
  $\epsilon_{\rm lim} = 1.1$.  Left panel: GA2; right panel: GA1.}
\label{fig:instability}
\end{figure}

To investigate the physical cause of this bar instability, we
consider two criteria commonly adopted in literature to
assess when a disk is unstable to bar formation.  The first one,
proposed by \cite{Efstathiou_1982} and based on N-body simulations,
predicts that the regulator of bar instability is the relative
contribution of the disk to the rotation curve:

\begin{equation}
 \Upsilon_{\rm disk} = \frac{V_{\rm disk}}{\sqrt{ GM_{\rm disk}/R_{\rm disk} } }\, ,
\end{equation}

\noindent
where, $M_{\rm disk}$, $R_{\rm disk}$ and $V_{\rm disk}$ are disk
mass, radius and velocity respectively.  Bar instability takes place
whenever $\Upsilon_{\rm disk} \leq \epsilon_{\rm lim}\simeq 1$.  To calculate
this quantity, we use for $M_{\rm disk}$ the mass of galaxy particles
within 10 kpc and with circularities $0.7<\epsilon<1.3$,
for $R_{\rm disk}$ the half-mass radius of the same particles, 
and for $V_{\rm disk}$ the maximum of the galaxy rotation curve within 10 kpc.
Results do not change if we use a different aperture to define disk mass and radius.
Figure~\ref{fig:instability} shows the quantity $\Upsilon_{\rm disk}$ as a
function of redshift, in both cases computed either using only star particles (blue
lines) or all galaxy particles (red line).  As usual, the left panel gives 
results for GA2, the right panel for GA1.  The shaded
regions denote values lower than a threshold $\epsilon_{\rm
  lim}=1.1$ \citep{Efstathiou_1982} applying to a pure stellar
disk, while a lower threshold of 0.9 has been suggested to apply to
gas disks. As a result, although $\Upsilon$
decreases with time, it takes on values in any case above
the suggested threshold.

The local stability of a self-gravitating disk to radial perturbations is usually
evaluated using the \cite{Toomre_1964} stability criterion. 
Though Toomre-unstable disks are expected to fragment into self-bound knots, 
a mildly unstable disk may develop a bar \citep{Julian66}. Furthermore
\cite{Athanassoula86} proposed
that $Q>2$ might be a general criterion against bar formation, since for
these high $Q$ values collective density waves become very weak and growth
rates of all instabilities are reduced.

To compute the Toomre parameter of a two component disk (with stars
and gas) we use, as in \cite{Monaco12}, the simplified approach of
\cite{Wang94}, that with high velocity dispersion
approximates well the more accurate expression
recently proposed by \cite{Romeo11}, which is our case.
We start from the separated component $Q$
values:

\begin{eqnarray}
Q_{*}(r) &=& \frac{\kappa \sigma_{r}}{3.36 G \Sigma_{*}} \\
Q_{g}(r) &=& \frac{\kappa \sigma_{r}}{3.36 G \Sigma_{g}} 
\end{eqnarray}

\noindent
where, for each component, $\Sigma(r)$ is its surface density,
$\sigma_r(r)$ its radial velocity dispersion and $\kappa(r) =
V(r)\sqrt{2+2d\ln V/d\ln r}/r$ the epicyclic frequency of the disk.
The effective $Q_{\rm tot}(r)$ parameter of the disk is computed as:

\begin{equation}
 Q_{\rm tot}(r) \simeq \left(\frac{1}{Q_{g}} + \frac{1}{Q_{*}} \right)^{-1}
\end{equation}

\noindent
Radial velocity dispersion is computed with the $3-\sigma$ rejection
method described in Section~\ref{section:vertical}, using as stellar
surface density the one obtained rejecting $>3-\sigma$ interlopers
\footnote{
We report here that, with respect to the Gaussian fit, the $3-\sigma$ rejection method
yields at the same time lower $\sigma_r$ and lower $\Sigma_\star$
(Figure~\ref{fig:dispersion}) and these corrections compensate in the
value of $Q_\star$, that is then very insensitive to the method used.
}.
In all cases, we find that gas has a minor
impact on $Q_{\rm tot}$. In fact, using $Q_*$ alone would lead to the same
conclusions.

\begin{figure}
\centering{
  \includegraphics[angle=90,width=\linewidth]{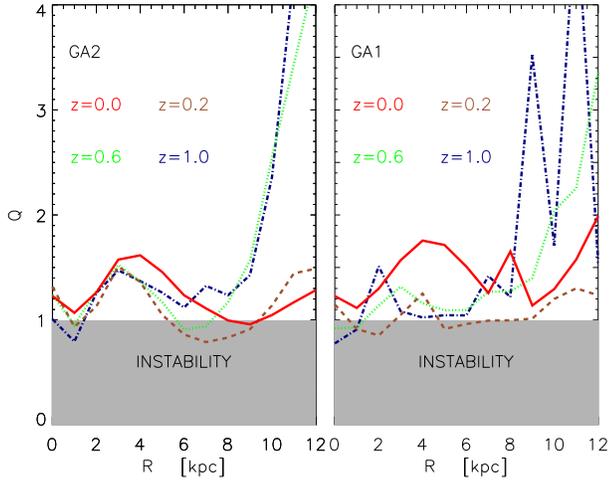}
  }
\caption{Toomre parameter $Q_{tot}(r)$ at $z = 0, 0.2, 0.6, 1.0$ (continuous red,
  dashed brown, dotted green and dot-dashed blue lines respectively).  The shaded areas show
  the region where disk are formally unstable. Left panel: GA2; 
  right panel: GA1.}
\label{fig:toomre}
\end{figure}

\begin{figure}
\centering{
  \includegraphics[angle=90,width=\linewidth]{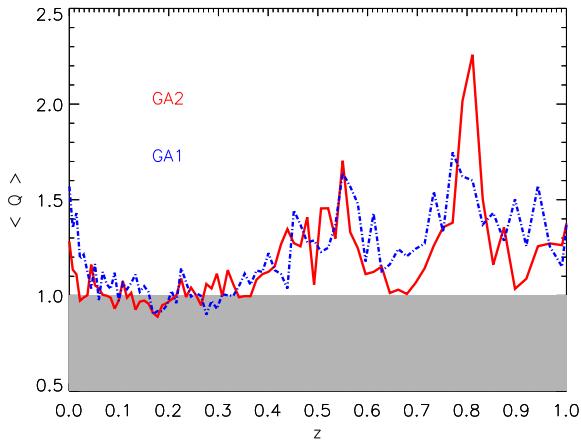}
  }
\caption{Average value of the Toomre parameter from 3 to 8 kpc, as a
  function of redshift, for GA1 (dot-dashed blue line) and GA2 (continuous red line).  The
  shaded areas show the region where disks are formally unstable.}
\label{fig:toomre_z}
\end{figure}

Figure~\ref{fig:toomre} shows $Q_{\rm tot}$ as a function of polar
radius at redshift 0, 0.2, 0.6 and 1.0.  Figure~\ref{fig:toomre_z} shows $\langle Q \rangle$,
the average value of $Q_{\rm tot}$ in the scale range from 3 to 8 kpc,
computed for all available outputs of the two simulations at $z\le1$.
In these plots the gray region denotes the $Q<1$ values of the Toomre
parameter where the disk is expected to be unstable.  
These galaxies are found to be formally stable at $z>0.3$, but
the Toomre parameter steadily decreases with time.  From
Figure~\ref{fig:toomre} we see that, at $z=0.6$, the disk of GA2 gets
weakly unstable both at the center and at $\sim6$ kpc.  As $z=0.2$
this second instability region has got wider, while $\langle Q\rangle$ has got below the value of unity since
$z\sim0.3$.  The GA1 galaxy becomes unstable at $r\sim2$ and $6$ kpc
only at $z=0.2$, but the evolution of  $\langle Q\rangle$ is
very similar to that of GA2.  For both galaxies, the rise of the
Toomre parameter at later time is due to the raise of $\sigma_r$, 
that is driven by the development of the radial streaming pattern.

Although the detailed behaviour of these galaxies is far from simple,
the start of instability roughly coincides with the time when $\langle
Q\rangle$ gets lower than the canonical threshold value of 1. 
Hence the
behaviour of bar instability in these galaxies is consistent with the
simple hypothesis that a bar is triggered by a secular instability.
Assuming the validity of the criterion $\langle Q\rangle<1$, the tidal
disturbance at $z=0.55$ takes place when the disk is still stable,
while disk heating due to tidal interaction leads to an increase
in velocity dispersion and therefore in $Q_{\rm tot}$ value. This may
explain why the bar-like feature of GA2 is transient.

\begin{figure}
\centering{
  \includegraphics[angle=90,width=\linewidth]{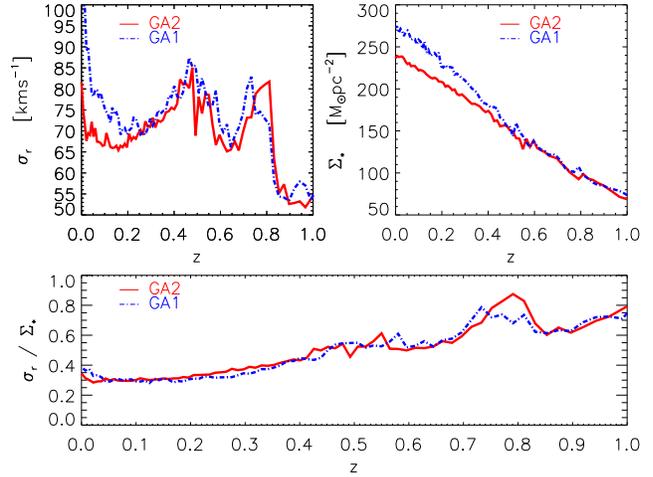}
  }
\caption{Disk radial velocity dispersion (top left panel), stellar disk
  surface density (top right panel) and their ratio as a function of
  redshift (bottom panel). All these quantities are averaged from 3 to
  8 kpc. Continuous red line for GA2 and dot-dashed blue line for GA1. }
\label{fig:sigmavz}
\end{figure}

The reason for the late-time secular decrease of $Q_{\rm tot}$ towards
the instability region is mostly due to the accumulation of the disk mass
rather than to a variation of disk velocity dispersion.  In
Figure~\ref{fig:sigmavz} we show the average disk radial velocity
dispersion, stellar surface density and the ratio of the two. All
quantities are evaluated in the range $3 \leq r \leq 8$ kpc and as a
function of redshift.  The radial velocity dispersion (top left of
Figure~\ref{fig:sigmavz}) is boosted by the tidal interaction with the
satellite and then it decreases before growing at $z \simeq 0.1$ in
both GA1 and GA2, while stellar surface densities (top right of
Figure~\ref{fig:sigmavz}) increase always with time.  Accordingly, their ratio
decreases with time (bottom panel of Figure~\ref{fig:sigmavz}).  This
demonstrates that the decrease of $Q_{\rm tot}$, that is the most
likely cause of the bar, is the accumulation of the disk mass at low
redshift, due to the continuous infall of gas into the DM halo.

\subsection{The role of halo triaxiality}
\label{section:triaxiality}

As mentioned in the introduction, halo triaxiality is a potential
trigger of bar instability, though the precise role of triaxiality has
been debated in the papers discussed above.  In those papers, halos
were extracted from collisionless N-body simulations and disks were
placed inside them.  As a matter of fact, this implies that the
gravitational influence that the formation of the disk has had on the
structure of the halo itself is neglected.
The impact of the formation of a
gaseous disk on the shape of a dark matter halo and its transformation
from prolate to oblate in the inner part was already studied in
isolated systems by \cite{Dubinski94}, \cite{Debattista_2008},
in cosmological environment
by \cite{Kazantzidis04}, \cite{Tissera10}, \cite{Abadi10} and more
recently by \cite{Zemp12}, \cite{Bryan_2013}.  
\cite{DeBuhr12} inserted live stellar
disks inside Milky Way-like dark matter halos from the Aquarius
simulations, finding a strong effect on the shapes of the inner halos
which evolve to become oblate.

In our paper I we have shown that, consistently with many other papers
\cite[e.g.][]{Governato12}, the inner slope of the dark matter halo of
our GA2 simulation is flatter than the typical $\rho\propto r^{-1}$
slope obtained when particles are collisionless. This is due to the
combined action of the adiabatic contraction caused by the formation of the
galaxy and the violent expansion due to episodic massive outflows. The
same process induces changes in the DM distribution in the region
occupied by the galaxy.

To address the influence that the formation of the galaxy has on
  the inner regions of the DM halo, we performed a simulation of the GA2
  switching off hydrodynamics, cooling and star formation, thus treating
  both DM and gas particles as collisionless particles; 
  we will call it GA2-cless in the following.
For the GA2 and GA2-cless simulations
we compute the inertia tensor of all DM particles
within a distance of $r_{200}/10$ (30.64 kpc for GA2-cless, 29.98 kpc
for GA2) 
and quantify the ratios among the
eigenvalues $I_i$ (with $i=1,2,3$), ranked in decreasing order.  In
GA2-cless we find $I_1/I_2=1.07$, $I_1/I_3=1.39$
and $I_2/I_3=1.30$, indicating a roughly prolate shape with
significant triaxiality.  In GA2, where the DM halo has hosted a
forming spiral galaxy, the ratios of
eigenvalues are $I_1/I_2=1.13$, $I_1/I_3=1.14$ and $I_2/I_3=1.01$,
indicating an oblate and nearly axisymmetric shape. Moreover, the
eigenvector corresponding to the largest eigenvalue of the inertia
tensor, i.e. to the direction where the halo is flattened, is found to be
aligned with the galaxy angular momentum within $4.37^\circ$.  This
alignment allows to infer that the oblate shape is due to the
formation of the disk itself.

We conclude that the influence of the triaxiality on disk dynamics should
not be addressed without considering, at the same time, the influence
that astrophysical processes bringing to disk formation have on the
triaxiality of the inner part of the DM halo.

\section{The Aq5 and Aq6 galaxies}
\label{section:AqC}

\begin{figure*}
\centering{
  \includegraphics[angle=90,width=\linewidth]{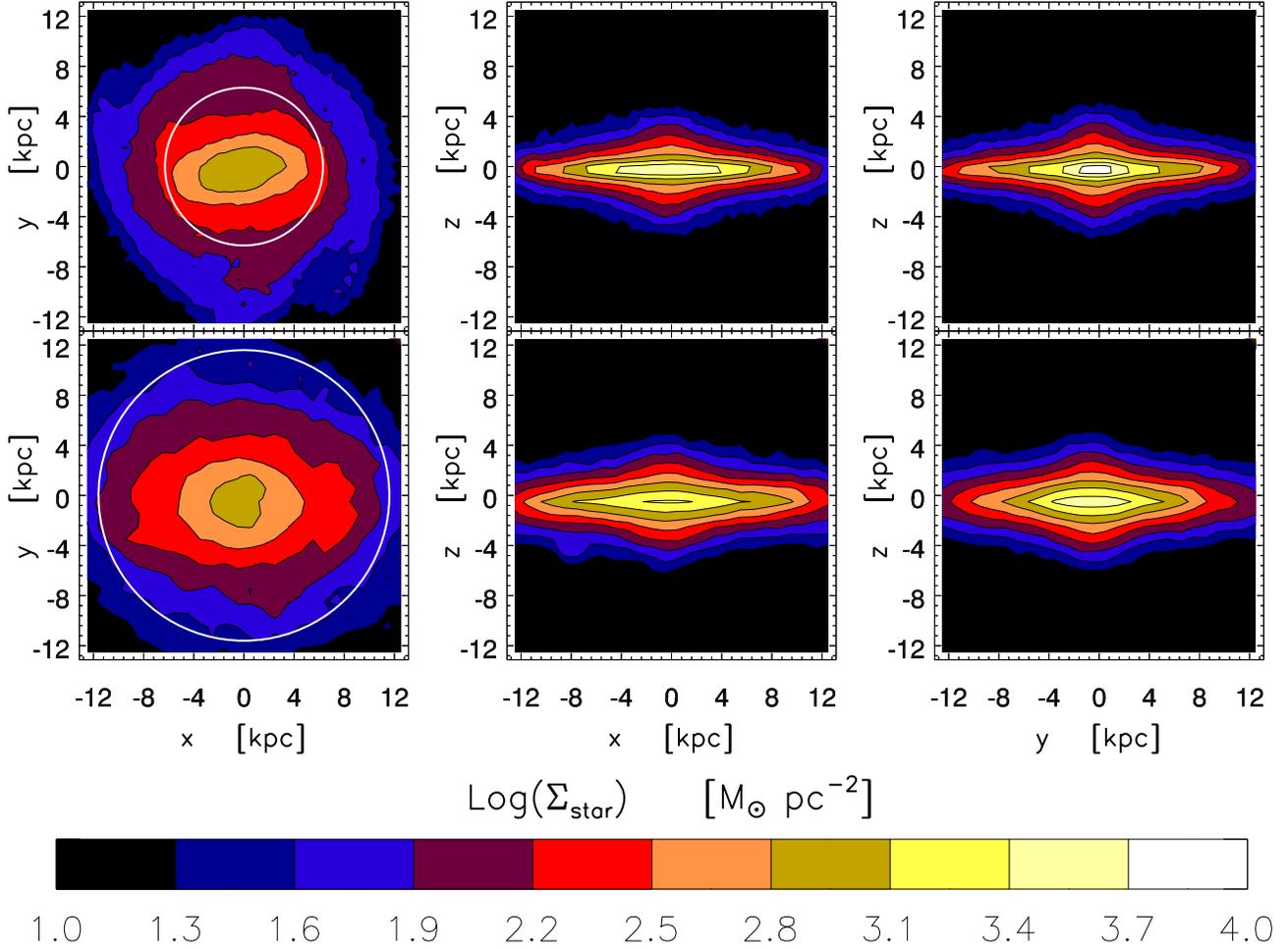}
  }
\caption{As in Figure~\ref{fig:maps}:
projected stellar maps of the AqC5 (top panels) and AqC6 (bottom
panels) galaxies at $z=0$. The white circle marks the bar length $L_{bar}$.}
\label{fig:maps_AqC}
\end{figure*}

\begin{figure}
\centering{
  \includegraphics[angle=90,width=\linewidth]{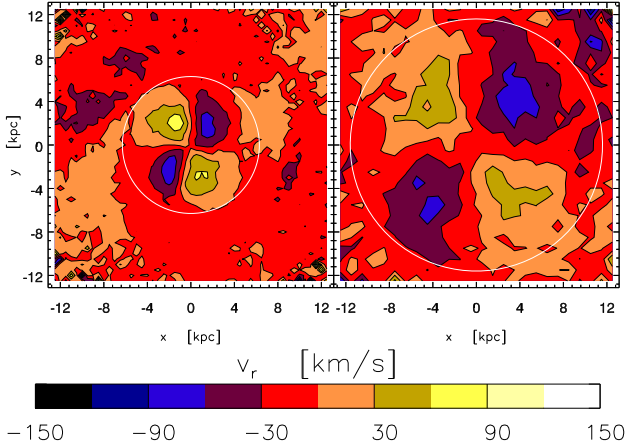}
  }
\caption{As in Figure~\ref{fig:streaming_motion}:
radial velocity maps on edge-on views of
  the two simulated AqC5 (left panel) and AqC6 (right panel) galaxies.
  The white circle marks the bar length $L_{bar}$.}
\label{fig:vel_r_AqC}
\end{figure}

\begin{figure}
\centering{
  \includegraphics[angle=90,width=\linewidth]{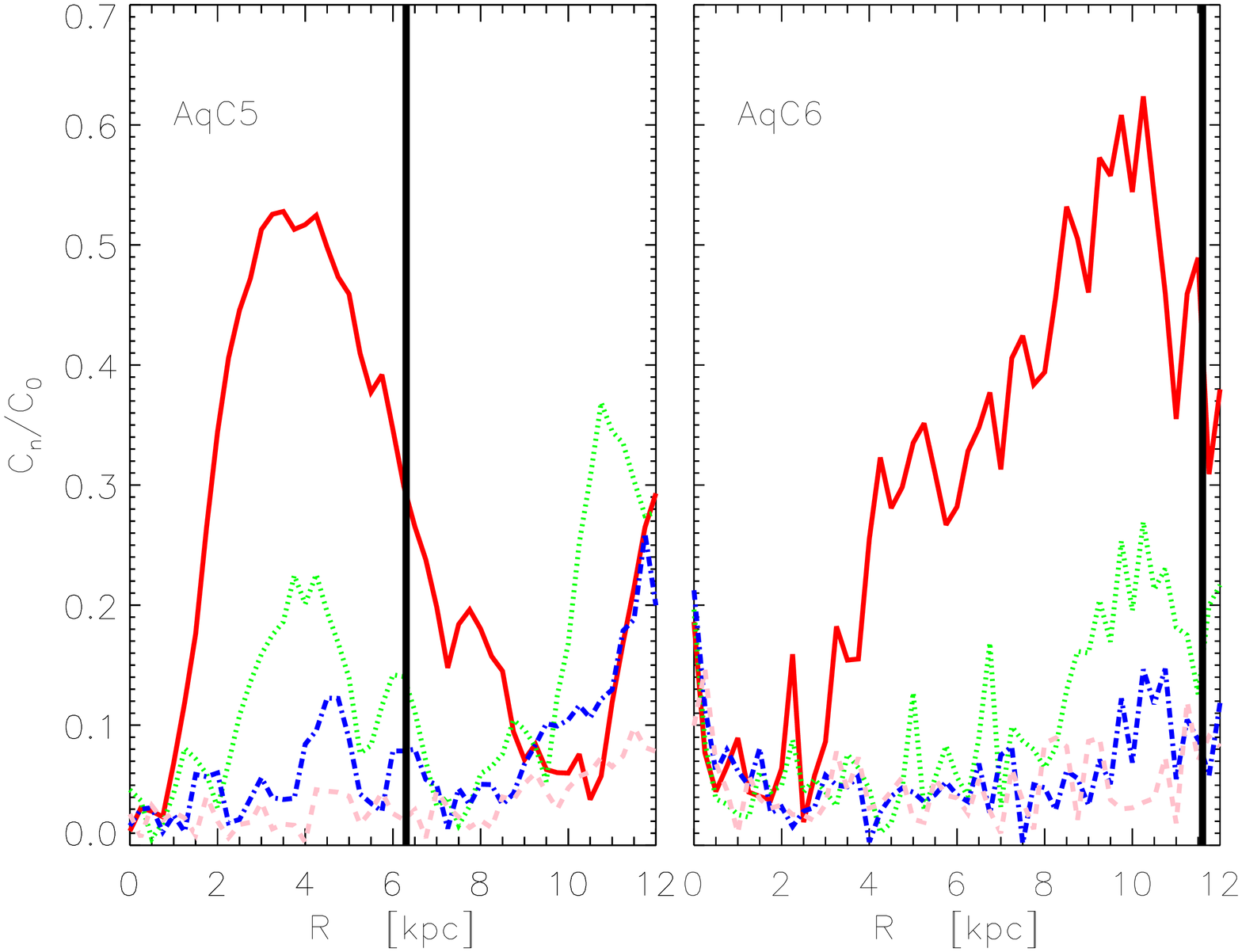}
  \includegraphics[angle=90,width=\linewidth]{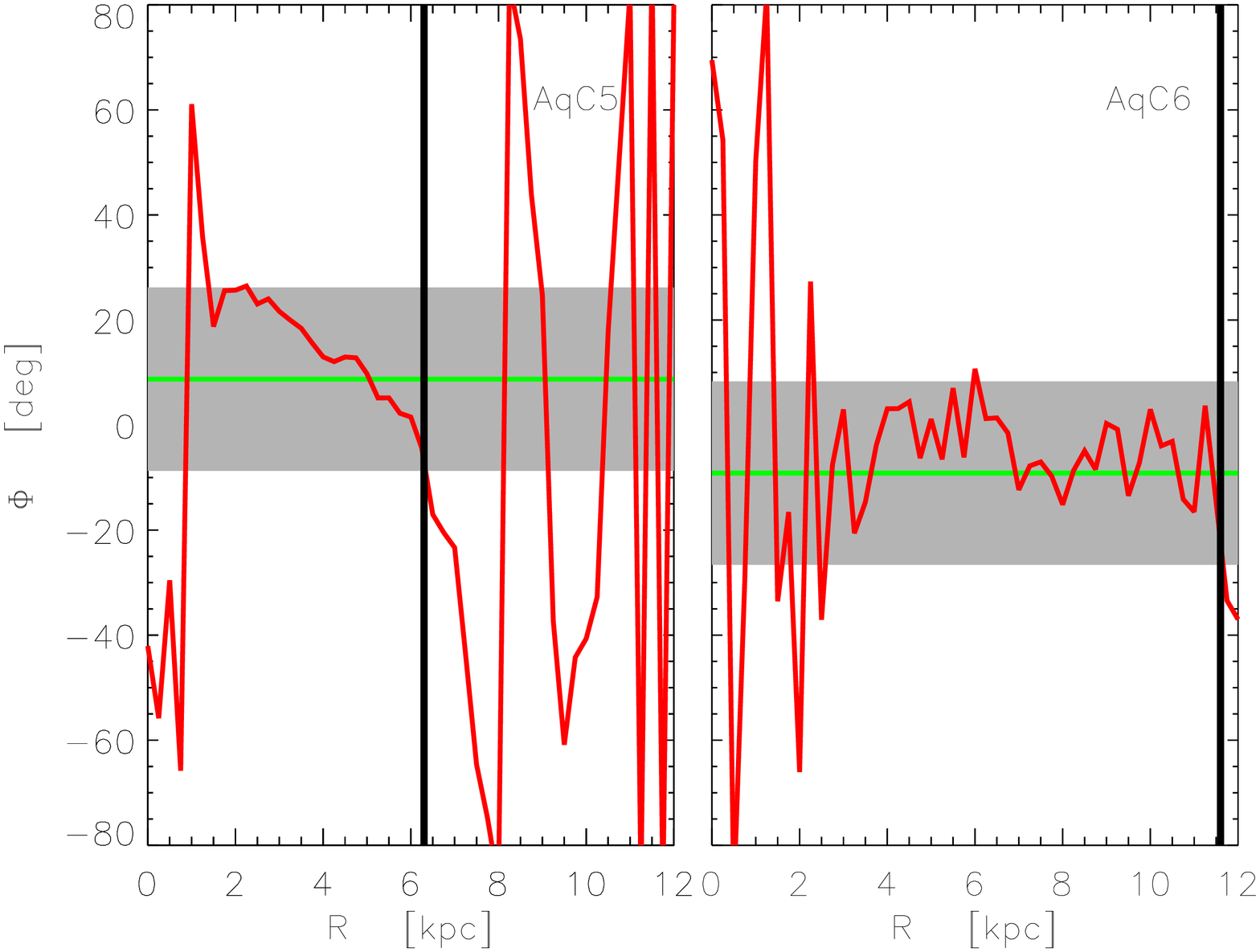}
  }
\caption{As in Figure~\ref{fig:fourier} and \ref{fig:phase}: Fourier
  analysis of the AqC5 (left panels) and AqC6 (right panels) galaxies.
  Upper panels show the amplitude of even Fourier coefficients, the
  lower panels the phase of $C_2$.  The black lines mark the bar
  length $L_{bar}$.}
\label{fig:fourier_AqC}
\end{figure}

\begin{figure}
\centering{
  \includegraphics[angle=90,width=\linewidth]{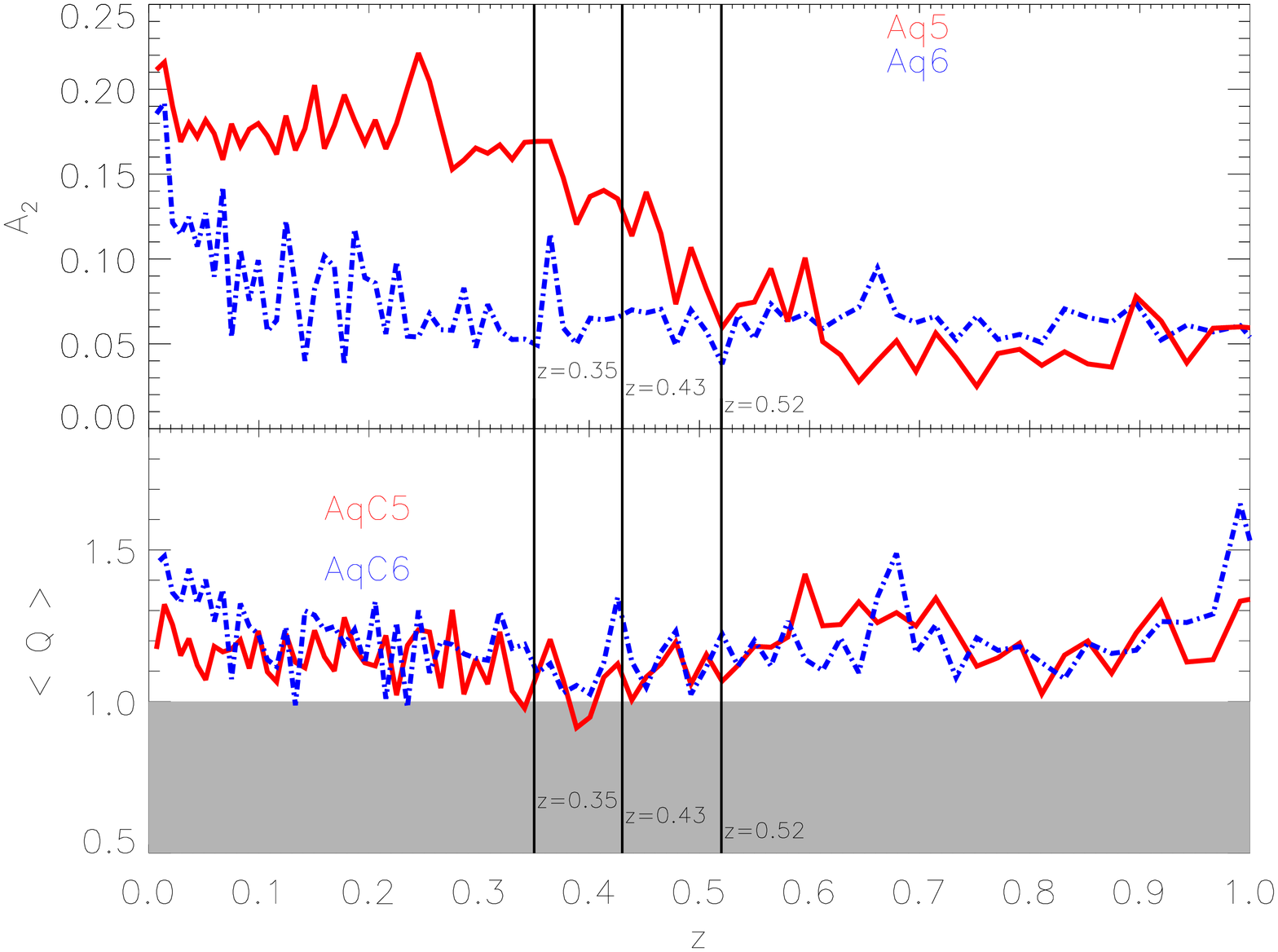}
  }
\caption{As in Figures~\ref{fig:amplitude} and \ref{fig:toomre_z}:
evolution with redshift of $A_2$ (upper panel) and $\langle Q\rangle$
(lower panel) for the AqC5 (continuous red lines) and AqC6 (dot-dashed blue lines).
Vertical lines give the times of near passages and final merger of the satellite.}
\label{fig:AqC_z}
\end{figure}

The two AqC simulations show a different behaviour with respect to the
GA ones.  Here we show the surface
density maps (Figure~\ref{fig:maps_AqC}), the face-on map of radial
velocities (Figure~\ref{fig:vel_r_AqC}), the Fourier coefficients and
phases (Figure~\ref{fig:fourier_AqC}) and the evolution of $A_2$ and
$\langle Q\rangle$, the latter being averaged again on $3<r<8$ kpc (Figure~\ref{fig:AqC_z}).  
The higher resolution AqC5 galaxy at $z=0$ has a bar with
properties similar to the one of GA2 (Figure~\ref{fig:maps_AqC}),
visible as an elongation of the isodensity contours, but with
a slightly smaller size with respect to the GA galaxies
(Figure~\ref{fig:maps_AqC}).  The signature of streaming motions very
clearly confirms the visual impression (Figure~\ref{fig:vel_r_AqC}).
The Fourier analysis (Figure~\ref{fig:fourier_AqC})
confirms the presence of a bar with maximal amplitude of $C_2/C_0
\sim0.5$ at $\sim4$ kpc and a bar length of $L_{\rm bar}=6.3$ kpc.
Spiral arms here give a much smaller contribution to the amplitude of
the $m=2$ mode.

The origin of this bar is however different: in Figure~\ref{fig:AqC_z}
we see that the $A_2$ coefficient starts to increase at $z\sim0.6$,
while the disk is stable according to $\langle Q\rangle$.  As
mentioned in Section~\ref{section:simulations}, this galaxy happens to
suffer a minor merger at roughly the same time as the GA galaxy.  In
particular, at $z=0.52$, the galaxy suffers a near passage at $\sim20$
kpc of a $1.6\times 10^9$ {\msun} satellite, with a mass ratio of
$1:32$ with respect to the main galaxy.  Further near passages are at
$z=0.43$ and $z=0.35$, when the satellite gets tidally destroyed.
These times are reported in Figure~\ref{fig:AqC_z} as vertical black
lines.  This coincides with the time interval where the $A_2$
coefficient increases from $\sim0.05$ to $\sim0.17$ and it supports the
idea that this bar is triggered by a tidal interaction.  The main
difference with respect to the GA galaxy lies in the higher mass
ratio, though the merger is still considered minor.

The AqC6 galaxy behaves differently.  Analogously to the GA
simulations, the perturbation from the satellite does not trigger a
bar.  Indeed, the dynamics of the merger is different in this case:
the satellite suffers a much slower orbital decay, so that the
apocenter of the orbit is still $\sim30$ kpc at $z=0.45$. The nearest
encounter, at $19$ kpc, is found only at $z=0.24$, while the tidal
disruption takes place at $z=0.14$.  At such late times the $Q_{\rm
  tot}$ parameter is close to 1 and the $A_2$ coefficient starts
to raise again as in the GA case.  While the cause in this case seems
to be the disk secular evolution (driven by the progressive accumulation
of the disk mass, as in Figure~\ref{fig:sigmavz}), we cannot exclude that
the bar is tidally triggered.

Figures~\ref{fig:maps_AqC}, \ref{fig:vel_r_AqC} and
\ref{fig:fourier_AqC} show that in this case the bar-like signatures
in streaming motions and Fourier analysis are relatively strong, but the bar
is very long, with an estimated length of $L_{\rm bar}=11$ kpc and
the isodensity contours are not very flattened.  The galaxy appears to
be caught during the early development of a very strong and long bar.

\section{Conclusions}
\label{section:conclusions}

In paper I we showed results of simulations of disk galaxies in
$\sim2-3\times10^{12}$ {\msun} halos, using two sets of initial
conditions (GA, \citet{Stoehr02} and AqC, \citet{Springel2008}, 
\citet{Scannapieco_2009})
at two resolutions.  In all cases, extended disks were obtained with low
$B/T$ ratios and general properties (disk size, mass surface density,
rotation velocity, gas fraction) that are consistent with observations
of the local universe.  These galaxies develop a bar at low redshift,
which is still consistent with the relatively high fraction ($\sim60$
per cent) of bars found in such massive spiral galaxies.  In this
paper we quantified the properties of these bars, starting from
morphology and kinematics of the inner region of the galaxies, then
performing a Fourier analysis of the mass surface density map
to assess the strength and length of the bars.  
We investigated
the physical conditions that cause bar instability and found that a
combination of low values of Toomre parameters and minor mergers can
explain the emergence of bars in our simulations.

Our main conclusions are the following:

(i) The close similarity of bar properties at $z=0$ and of the
  development of the instability at $z\le0.3$ for GA2 and GA1, which
  differ in mass resolution by a factor of 9.3, disfavors
  the hypothesis that these bars are a result of a numerical
  instability due to poor resolution.

(ii) Despite the softening does not allow to properly resolve the
vertical structure of the disk, we find our simulated disks to have
vertical and radial velocity dispersion compatible with observations,
at least for $r<10$ kpc.

(iii) In the GA simulations the morphology and kinematics of the inner
$\sim10$ kpc are fully consistent with the presence of a bar.
In both cases the typical kinematic
signatures of increased line-of-sight velocity dispersion and radial
streaming motions are present.  The Fourier analysis shows that $C_2/C_0$
peaks to values of 0.4 and 0.7 in the GA2 and GA1 cases, that the
bar length is $L_{\rm bar}\simeq 8.8$ kpc in both cases.

(iv) The time evolution of $A_2$ shows that this instability starts at
$z=0.2$ and quickly rises at $z=0$ for both GA2 and GA1, so the
difference in bar strength is likely due to a small time offset in
the growth of the structure.

(v) Before the onset of bar instability, disks result to be stable
according to the criterion proposed by \cite{Efstathiou_1982} and
they are very close to the threshold of Toomre instability, with $Q_{\rm tot} \simeq 1$.

(vi) A minor merger taking place from $z=0.57$ to $z=0.35$, with a
stellar mass ratio of $1:50$, results, especially in GA2, in a
transient tidal disturbance with high values of the $A_2$ coefficient
which disappears as quickly as it has appeared. A Fourier analysis
reveals that the perturbation is due to a strong spiral pattern rather
than to a bar.

(vii) The AqC simulations follow a different path.  At a higher
resolution (AqC5), a minor merger taking place from $z=0.52$ to
$z=0.35$ induces a true bar that lasts until the end of the
simulation.  The $\langle Q\rangle$ parameter is greater than 1 before the first
close encounter with the satellite, but the mass ratio is higher in
this case ($1:32$) and this likely justifies the different behaviour of
AqC5 with respect to GA2.  The final bar has a length of $L_{\rm bar}=6.5$ kpc.

(viii) The minor merger in AqC6 takes place at later times.  In this
case the disk gets barred in a way similar to the GA galaxy.  
However, in this case the role of the merger in the triggering of the bar cannot be excluded.
The resulting bar is very long ($L_{\rm bar}=11$ kpc) and it is caught in
a relatively early phase of development.

(ix) We find that the formation of the disk influences the triaxiality
of the inner regions of the DM halo. 
In the GA2-cless simulation, performed treating both DM and gas particles as collisionless particles,
the DM halo is triaxial, while in GA2, where the halo has hosted
the formation of a spiral galaxy,
the inertia tensor of the DM halo
is roughly oblate and its eigenvector corresponding to the largest eigenvalue is
found to be well aligned with the galaxy angular momentum.
Thus, special care is needed in addressing the role of triaxiality on bar instability 
when disks are embedded in DM halos extracted from collisionless simulations.

Overall, our simulations are consistent with a relatively simple
picture of bar instability being triggered either by secular processes
in Toomre-unstable disks ($Q_{\rm tot}\lesssim1$) or by minor mergers, when
the stellar mass ratio is at least of order $1:30$.  With mass ratios
as small as $1:50$, the merger can stimulate transient features that
may look like bars but do not pass a test based on Fourier phases.  Of
course, these simulations do not give sufficient statistics to provide
this picture with the proper justification.
We tested several versions
of our code on these sets of IC and we noticed that bars may or
may not come out, depending on the detailed state of the galaxy.  For
instance, the AqC5 galaxy in the \cite{Scannapieco12} paper, run with
a previous version of our code with pure thermal feedback, primordial
cooling and no chemical evolution, showed a very strong bar (the whole
disk had collapsed into a cigar-like structure).  A similar
thing happens to GA2 when simulated with the same code.  In this two
cases we found that the bar is triggered by the two minor mergers
discussed above, while the disks are Toomre-stable before the merger.
The difference in this case is that, due to the fact that feedback is
less effective in limiting star formation in small halos, mass ratios
are much larger.

As a concluding remark, our results are
found to be stable with resolution at
least in the GA case, hence we consider the origin of the bar physical
rather than purely numerical.
However, the presence of a bar
depends on the fine details of the disk structure and its environment that
are not yet numerically under full control, as the different timing of
the satellite merging in AqC5 and AqC6 testifies.  So these results are
not a premise to a robust prediction of the presence of bars in
simulated disks, but it will certainly allow to better understand the
emergence of this complex phenomenon that, although it has been observed 
since the beginning of extragalactic astronomy, is still not well
understood.

\section*{Acknowledgments}
We thank Volker Springel who provided us with the non-public version
of the GADGET-3 code. We acknowledge useful discussions with Stefano
Borgani, Gabriella De Lucia, Emiliano Munari and Marianna Annunziatella.  
The simulations were
carried out at the ``Centro Interuniversitario del Nord-Est per il
Calcolo Elettronico'' (CINECA, Bologna), with CPU time assigned under
University-of-Trieste/CINECA and ISCRA grants, and at the CASPUR
computing center with CPU time assigned under two standard
grants. This work is supported by the PRIN MIUR 2010-2011 grant ``The
dark Universe and the cosmic evolution of baryons: from current
surveys to Euclid'', by the PRIN-MIUR 2012 grant ``Evolution of Cosmic
Baryons'', by the PRIN-INAF 2012 grant ``The Universe in a Box:
Multi-scale Simulations of Cosmic Structures'', by the INFN ``INDARK''
grant, by the European Com- missions FP7 Marie Curie Initial Training
Network CosmoComp (PITN-GA-2009-238356), by the FRA2012 grant of
the University of Trieste and by ``Consorzio per la Fisica di Trieste''.

\bibliographystyle{mn2e}
\bibliography{master}

\label{lastpage}

\end{document}